\documentclass[12pt,oneside, a4paper]{article}

\pdfoutput=1  
 
\ifx\pdfoutput\undefined
\usepackage[dvips,bookmarks=false]{hyperref}	
\else
\usepackage{hyperref}	
\fi
\hypersetup{colorlinks,bookmarksopen,bookmarksnumbered,citecolor=blue,linkcolor=blue,pdfstartview=FitH,urlcolor=blue}

\usepackage{tikz}
\usepackage{amsfonts}

\usepackage{graphicx}
\usepackage{amssymb}
\usepackage{cite}
\usepackage{bm}
\usepackage{amsmath,amsthm}
\usepackage{cleveref}
\usepackage{csquotes}
\usepackage{siunitx}
\usepackage{slashed}
\usepackage{multirow}
\usepackage{multicol}
\usepackage{xcolor,colortbl}
\usepackage[dvipsnames]{xcolor}
\usepackage{booktabs}
\usepackage{subcaption}
\usepackage{tabularx}
\numberwithin{equation}{section}

\usepackage{tcolorbox}

\usepackage[toc,page]{appendix}
\definecolor{light-gray}{gray}{0.95}
\usepackage{tcolorbox}
\usepackage{bm}
\usepackage[normalem]{ulem}

\begin{document}

\begin{titlepage}
\begin{center}
        {\Large\bf Non-Holomorphic $A_4$ Modular Symmetry in Type-I Seesaw: Implications for Neutrino Masses and Leptogenesis} 
\vspace{1cm}

\renewcommand{\thefootnote}{\fnsymbol{footnote}}
{\bf Swaraj Kumar Nanda}$^a$\footnote[1]{swarajnanda.phy@gmail.com}
{\bf, Maibam Ricky Devi}$^{b}$\footnote[4]{deviricky@gmail.com}
{\bf, Sudhanwa Patra}$^{c,d}$\footnote[2]{sudhanwa@iitbhilai.ac.in}
\vspace{5mm}

$^a$ {\it {Department of Physics, ITER, Siksha ’O’ Anusandhan, Deemed to be University, Bhubaneswar-751030, India}}\\
$^b$ {\it {Department of Physics, Gauhati University, Guwahati 781014, India}}\\
$^c$ {\it {Department of Physics, Indian Institute of Technology Bhilai, Durg-491002, India}}\\
$^d$ {\it
Institute of Physics, Sachivalaya Marg, Bhubaneswar- 751005, India}

\vspace{4mm} 


\abstract{We propose a minimal extension of the Standard Model with right-handed neutrinos, governed by a non-holomorphic $A_{4}$ modular flavor symmetry of level $N=3$. Within this model framework, the light neutrino masses are generated via the popular type-I seesaw mechanism in which the structure of the Dirac neutrino Yukawa couplings is decided by nonholomorphic modular forms. Unlike conventional flavor models with ad hoc flavon fields, the structure of Dirac and Majorana mass matrices is entirely determined by a modulus parameter $\tau$. 
We construct the predictive mass matrices for charged leptons, Dirac neutrinos, and right-handed Majorana neutrinos and show the compatibility with neutrino oscillation data by an appropriate choice of input model parameters.  We find that our analysis of neutrino masses and mixing gives excellent agreement with current neutrino oscillation observables by taking normal hierarchical pattern of light neutrinos.   We present numerical analysis of two sets of benchmark points explaining neutrino masses while generating the correct amount of baryon asymmetry via thermal leptogenesis.  We estimate numerically the values of CP-asymmetry and examine the evolution of the lepton asymmetry by studying Boltzman equations by considering both strong and washout regimes with CP-asymmetry parameter in the range $|\varepsilon_{1}| \sim 10^{-4}$--$10^{-8}$. The model predicts an effective Majorana mass in the few meV range, below current experimental bounds but within reach of next-generation $0\nu\beta\beta$ searches.  The key feature of non-holomorphic $A_4$ modular symmetry naturally accommodates non-zero neutrino masses and mixings, minimizes the Yukawa arbitrariness, and establishes a direct connection between high-scale leptogenesis with low-energy neutrino observable parameters, thereby the model provides a testable link between neutrino flavor physics and cosmology. }
\end{center}

\end{titlepage}

\renewcommand{\thefootnote}{\arabic{footnote}}
\setcounter{footnote}{0}

\tableofcontents







\section{Introduction}
Over the past few decades or so, neutrino oscillation experiments~\cite{SNO:2002tuh, T2K:2019efw, DayaBay:2012fng, DoubleChooz:2011ymz} have provided conclusive evidence that neutrinos are massive and undergo flavor mixing. After a relentless row of research and development in neutrino experiments, physicists have now arrived at an era where they can understand neutrinos to a great extent and measure their neutrino oscillation parameters, such as the mass squared differences and the mixing angles, with unprecedented accuracy. The fact that neutrinos are massive and mix with each other gives compelling evidence for physics beyond the standard model as neutrino are massless in SM. Unlike the other fermions, we still do not have any information on the absolute masses of the neutrinos. This gets further complicated when the mixing pattern of the neutrinos seems unrelated to those of the quarks and charged leptons, foraying its mass-hierarchy~{\cite{SNO:2001kpb}, \cite{KamLAND:2002uet}, \cite{Super-Kamiokande:1998kpq}} and its nature - whether Dirac \cite{Dirac:1928hu} or Majorana type \cite{Majorana:1937vz}, questionable. 

Even after these spectacular findings, there are some fundamental questions at the the theoretical and experimental level that remained to be answered in order to understand neutrino physics. The well-determined values of neutrino mass-square differences and mixing angles motivated to propose a simpler extension of the SM to understand the origin and structure of neutrino masses and mixing. One of the elegant idea to explain non-zero neutrino masses and mixing is popularly known as seesaw mechanism, where heavy messenger degree of freedom couple to SM neutrino, can generate small masses for light active neutrinos naturally through a  suppression factor with a large messenger mass scale. The most minimal and well-studied realization is the type-I seesaw mechanism~\cite{Minkowski:1977sc, Mohapatra:1979ia, Yanagida:1979as, Gell-Mann:1979vob}, which introduces three SM-singlet right-handed neutrinos, allowing the light neutrino masses to emerge as the low-energy remnants of a high-scale Majorana mass matrix. Alternative seesaw mechanisms, including type-II~\cite{Magg:1980ut, Schechter:1980gr, Cheng:1980qt, Lazarides:1980nt, Mohapatra:1980yp} and type-III~\cite{Foot:1988aq,He:2012ub}, introduce scalar triplets and fermion triplets respectively, offering different model-building avenues. In one of our recent works \cite{Nanda:2025fvw}, we have introduced double-seesaw mechanism to study the neutrino phenomenology.

The observed baryon asymmetry quantified by the baryon-to-photon ratio of the universe is a fascinating topic that has motivated the scientific community in particle physics and cosmology to understand its dynamics for a long time. The observed baryon asymmetry of the Universe is $\eta_B \equiv n_B/n_\gamma \simeq (6.1\pm0.04)\times10^{-10}$~\cite{Planck:2018vyg,Cooke:2017cwo}. It is the most precisely observed from cosmic microwave background (CMB) measurements~\cite{Cooke:2017cwo} and independently, by Big-Bang nucleosynthesis (BBN) determinations of primordial deuterium~\cite{Cooke:2017cwo}. It is already understood that the observed baryon asymmetry of the universe can not be addressed within the SM, thereby hinting towards possible new high-scale dynamics in the early universe. Alternatively, leptogenesis, initially proposed by Fukugita~\cite{Fukugita:1986hr}, has been a widely studied phenomenon in early universe cosmology that remains one of the compelling and minimal frameworks that links neutrino mass to baryogenesis explaining the observed baryon asymmetry of the universe~\cite{Davidson:2008bu, Langacker:1986rj, Luty:1992un, Murayama:1992ua, Buchmuller:1996pa, Covi:1996wh, Sakharov:1967dj}. Leptogenesis is concieved with the idea that a net lepton asymmetry is created via the CP-violating and lepton number violating, out of equilibrium decays of heavy right-handed Majorana neutrinos and later, the created lepton asymmetry is processed to a baryon asymmetry by the anomalous $B+L$ violating sphaleron processes~\cite{Klinkhamer:1984di, Arnold:1987mh, Kuzmin:1985mm, Rubakov:1996vz}. 
The CP asymmetry $\varepsilon_i$ involved in leptogenesis depends on combinations of Dirac Yukawa couplings $\mbox{Im}[(Y_D^\dagger Y_D)^2_{21}]$ and on the heavy neutrino mass spectrum ($M_i$). This implies that different realizations of the seesaw mechanisms (type-I, type-II, type-III, and their minimal variants) generically produce very different high-scale CP-violating effects even when they reproduce the same low-energy light neutrino masses. Therefore, exploring minimal SM extensions with fully determined or tightly constrained flavour structures -- those emerging from discrete or modular symmetries -- is especially appealing by reducing the number of free parameters. Such minimal frameworks explains light neutrino masses consistent with neutrino oscillation data and makes leptogenesis predictive allowing a direct connection between low energy neutrino observables ($0\nu\beta\beta$, cosmological mass bounds) and matter-antimatter asymmetry of the universe. 

A longstanding puzzle in flavor physics lies in the observed pattern of
fermion masses and mixings, often referred to as the flavor problem. 
Various  ansatzes have been proposed to study these interesting flavor structuring discrete symmetries such as $A_4$, $S_4$, $T^\prime$, and $Z_N$ etc. to explain nonzero neutrino masses and mixing in agreement with oscillation data~\cite{Petcov:2017ggy, Ding:2024ozt, Ishimori:2010au,Chauhan:2023faf}. The main goal behind most of flavor models is to derive a simple and predictive structure of the Yukawa couplings, thereby addressing the known hierarchies and mixing angles in the neutrino sector. Of all recently explored flavor symmetries, the $A_4$ symmetry is the most widely used, as it can naturally accommodate the left-handed triplet leptons and the three right-handed singlet leptons. However, these flavor-motivated framework requires extra scalar fields called as flavons, to break the symmetry appropriately and ensure the vacuum expectation values (VEVs) alignment correctly. The disadvantage with these flavor-rich models is that the scalar potential becomes complicated and required fine-tuned VEV allignment for producing correct fermion masses and mixing.

A significant breakthrough occurred with the original proposal by Feruglio in 2017~\cite{Feruglio:2017spp} and followed by \cite{Ferrara:1989bc, Ferrara:1989qb, Feruglio:2019ybq, Ohki:2020bpo}, where it was shown that modular symmetries, originally arising from string theory compactifications, could serve as a natural organizing principle for flavor structures and neutrino mass model building. Thus, modular symmetry garnered attention, as its incorporation in a neutrino model seems a promising approach to solve this flavor problem and hierarchical mass patterns naturally without the inclusion of excessive flavon fields. Unlike neutrino models where additional Higgs-like fields called \enquote{Flavons} are used to construct a viable model, the neutrino modeling process is simplified to a large extent by incorporating Yukawa couplings as modular forms instead of assigning additional symmetries like discrete cyclic groups, $ Z_{n} $. This simplified the model building aspects considerably and gives predictive power with small sets of input model parameters. Among the already explored modular groups, $\Gamma_3 \simeq A_4$ is widely explored and has received much attention due to its minimality and suitability for explaining realistic lepton masses and mixing patterns. Additionally, supersymmetric (SUSY) version of modular symmetry has been investigated with the requirement of superpotential to be holomorphic. Even though model building in SUSY frameworks are elegant but the strict holomorphicity condition imposes constraints on modular space and thus, limiting its phenemelogical applications in neutrino sector. Some recent contributions on modular symmetry based neutrino models at SUSY framework can be found in \cite{Chen:2024otk, Ding:2024pix, Nomura:2019xsb, RickyDevi:2024ijc, Devi:2023vpe, Singh:2024imk, Kashav:2025mch, Kashav:2024lkr, Kashav:2021zir, Behera:2025tpj, Mishra:2024vhj, Mishra:2024fcr, Behera:2024hya, Behera:2024ark, Behera:2024vfv, Kumar:2023moh, Mishra:2023ekx, Mishra:2023cjc, Mishra:2022egy, Behera:2020lpd, Granelli:2025lds, Marrone:2025shp, Petcov:2024vph, Chen:2025tby, Arriaga-Osante:2025ppz,Ding:2023ydy, Knapp-Perez:2023nty, Liu:2021gwa, Qu:2021jdy, Ding:2021iqp, Yao:2020zml, Ding:2020zxw, Liu:2020msy, Liu:2020akv, Lu:2019vgm, Ding:2019gof, Liu:2019khw, Ding:2019zxk, Ding:2019xna }.

Alternatively, non-supersymmetric (non-SUSY) modular theories with non-holomorphic modular forms allow flexibility in the Yukawa sector and thus opening new avenues for studying neutrino masses, lepton flavor violation, dark matter, baryon asymmetry of the universe etc. The prime difference between modular symmetry and conventional flavor symmetry models is that in case of modular symmetry, the Yukawa couplings depend on the complex modulus parameter $\tau$ and they transform nontrivially under the transformation rules of modular symmetry. The most explored modular groups $\Gamma_{N}$ such as $A_4$, $S_4$, $A_5$ etc can predict excellent correlations between neutrino oscillation parameters hinting for possible new physics.  

The interesting idea of non-holomorphic formalism in neutrino physics model building was originally introduced by Qu, Ding in their work~\cite{Qu:2024rns} in July 2024 to formulate neutrino mass models within non-supersymmetric framework. Many attempts have already been made to investigate series of neutrino mass models in non-SUSY extension of SM by using non-holomorphic modular forms and various seesaw mechanisms to analyze different aspects of neutrino phenomenology
~\cite{Nomura:2024atp, Nomura:2024vzw, Nomura:2024nwh, 
Nomura:2025ovm, Nomura:2025raf, Nomura:2025bph, 
Kang:2024jnp, Ding:2024inn, Li:2024svh, Okada:2025jjo, Kobayashi:2025hnc, Loualidi:2025tgw, Zhang:2025dsa, Priya:2025wdm, Nomura:2024ctl, Abbas:2025nlv, Li:2025kcr, Dey:2025zld}. In the context of type-I seesaw mechanism, Ding, Lu, Petcov and Qu \cite{Ding:2024inn} proposed $A_4$ modular symmetry to construct Dirac neutrino and Majorana mass for right-handed neutrino while considering charged lepton already in diagonal basis but their analysis limited to neutrino masses and mixing analysis.    

In the present analysis, we investigate a predictive realization of this idea of modular $A_4$ symmetry in the type-I seesaw mechanism with simplified flavour structure of the Dirac neutrino mass matrix $M_D$ as well as for the Majorana masses for right-handed neutrinos. Following the pioneered work~\cite{Feruglio:2017spp} and subsequently,  non-holomorphic (non-SUSY) formulations~\cite{Qu:2024rns}, we implement the type-I seesaw where SM is minimally extended with right-handed neutrinos (RHNs) $N_{R_{i}}$ ($i=1,2,3$) and impose non-holomorphic modular forms of weights $k=0,\pm 2$ to derive the structure of the Yukawa multiplets $Y^{(k)}_{3}(\tau)$. Our construction differs from earlier modular $A_4$ studies in three important respects: (i) we allow a non-diagonal charged-lepton mass matrix so that \(U_{\rm PMNS}=U_\ell^\dagger U_\nu\) and charged-lepton mixing contributes non-trivially to low-energy observables, (ii) we work in a non-SUSY (polyharmonic Maass or non-holomorphic) modular  type-I framework that gives greater flexibility in the Yukawa couplings in the neutrino sector while retaining strong predictive power~\cite{Qu:2024rns, Ding:2024inn}, and (iii) we explicitly compute the high-scale CP asymmetries and solve the Boltzmann equations to assess whether the same modular parameter region that fits oscillation data can also produce the observed baryon asymmetry of the universe.   
By scanning the complex modulus parameter $\tau$ together with a small set of complex prefactors $\{\alpha_D, \beta_D, \gamma_D\}$ and RHN mass scales $\{\Lambda_{1,2}\}$, we identify regions where neutrino masses, mixing angles and the baryon asymmetry are simultaneously accommodated — thereby offering a directly testable bridge between low-energy neutrino data and early-Universe baryogenesis. 

The structure of the paper is as follows. We present the adequate theoretical formulation of modular $A_4$ symmetry including transformation properties of particles and Yukawa couplings and assignment of proper modular weight in Sec-2, In Sec-3, we discuss the methodology of detailed numerical analysis and parameter scanning. We show the phenomenological results for neutrino oscillation observables in Sect.4 and we discuss the role of modular $A_4$ symmetry in estimating CP-asymmetry and study the evolution of lepton asymmetry via the Boltzmann equations to explain correct baryon asymmetry of the universe via themral leptogensis. Finally, we conclude the summary of our findings in Sect.5. 

\section{Modular group \texorpdfstring{$\Gamma_{3} \equiv A_{4}$}, polyharmonic Maass forms and modular Yukawa couplings}
\label{sec:modular_background}
This section outlines the essential mathematical framework and notation underlying the construction of modular-invariant Yukawa couplings based on the finite modular group $\Gamma_{3}\simeq A_{4}$. Modular forms, described by Eichler as the “fifth fundamental operation” beyond the conventional arithmetic operations, play a crucial role in this formulation. Their name indicates their deep connection with the moduli space of complex curves, particularly genus-one Riemann surfaces \cite{Bost1992}. These functions encode rich symmetry properties on the upper half-plane, with their behavior often determined through
analysis within the fundamental domain. Classical modular objects are holomorphic (or meromorphic) functions, while almost non-holomorphic extensions allow for polynomial dependence on $1/\text{Im}(\tau)$ with holomorphic coefficients. For our purposes, we summarize the points:
\begin{itemize}
\item[(i)] the modular transformation properties of the complex modulus $\tau$,
\item[(ii)] the relevant classes of modular forms—including polyharmonic Maass forms—that enter the flavor construction, and
\item[(iii)] the modular invariance conditions restricting the permissible Yukawa structures in the Lagrangian.
\end{itemize}

A complex function $f(z)$ can generally be expressed as, $f(z)=u(x,y)+i\,v(x,y)$ where $z=x+i\, y$ and $i=\sqrt{-1}$. When the Cauchy–Riemann conditions are satisfied, $f(z)$ is holomorphic, otherwise, it is specified as non-holomorphic. The Wirtinger calculus \cite{wirtinger1927formalen} provides a convenient formalism to treat such functions, by introducing derivatives with respect to $z$ and its complex conjugate $z^{*}$. This framework is particularly useful for analyzing non-holomorphic modular objects. Notable examples include the non-holomorphic Eisenstein series and Maass forms, the latter being eigenfunctions of the hyperbolic Laplacian that play an important role in the theory of automorphic functions.

Let the complex modulus be denoted by
$
\tau = x + i y,\qquad x=\Re\tau,\; y=\Im\tau>0 $,
so that \(\tau\) lives in the upper half plane \(\mathcal{H}\). The full modular group \(SL(2,\mathbb{Z})\) acts on \(\tau\) by linear fractional transformations
\begin{equation}
\tau \longmapsto \gamma\tau \;=\; \frac{a\tau+b}{c\tau+d},\qquad
\gamma=\begin{pmatrix}a & b\\ c & d\end{pmatrix}\in SL(2,\mathbb{Z}),\quad ad-bc=1.
\label{eq:modaction}
\end{equation}
Finite level subgroups $\Gamma_{N}$ of $SL(2,\mathbb{Z})$ give rise to finite quotient structures that can be plotted to well-known non-Abelian discrete groups for small values of $N$. These identifications play a crucial role in connecting modular invariance with flavor symmetries in particle physics. For example, when $N=3$, the corresponding finite modular group is isomorphic to the alternating group of four elements, $\Gamma_{3} \simeq A_{4}$. This correlation provides a natural origin for the study of $A_{4}$ symmetry in flavor model building,particularly in explaining the observed patterns of lepton mixing.

Conventional modular forms are holomorphic functions \(f:\mathcal{H}\to\mathbb{C}\) that transform with a weight \(k\in\mathbb{Z}\)
\begin{equation}
f(\gamma\tau) = (c\tau+d)^k\, f(\tau)\qquad (\gamma\in SL(2,\mathbb{Z})),
\end{equation}
and that are holomorphic at the cusps. Relaxing holomorphicity leads to several useful generalizations \cite{Bruinier2002OnTG, bringmann2011mockperiodfunctionssesquiharmonic}:
\begin{itemize}
  \item \emph{Maass forms}: are real-analytic functions on \(\mathcal{H}\) that transform like modular forms but are eigenfunctions of the hyperbolic Laplacian (not required to be holomorphic).
  \item \emph{Harmonic Maass forms}:Maass forms with Laplacian eigenvalue vanishes (they are annihilated by the Laplacian operator) \cite{Bruinier2002OnTG}.
  \item \emph{Polyharmonic Maass forms (PMFs)}: generalizations wipe out by a finite repeat of the weight \(k\) Laplacian, thereby providing a broader framework for studying non-holomorphic modular objects with rich mathematical and physical applications. \cite{Lagarias_2016}
    \[
    \Delta_k^{\,d+1} f = 0,\qquad \Delta_k^{\,d} f \neq 0,
    \]
    where \(\Delta_k\) denotes the weight-\(k\) hyperbolic Laplacian and \(d\) is the depth.
\end{itemize}
The hyperbolic Laplacian of weight $k$ is defined as~ \cite{Bringmann2017HarmonicMF}
\begin{equation}
\Delta_{k} \;=\; -\,y^{2}\left(\frac{\partial^{2}}{\partial x^{2}} + \frac{\partial^{2}}{\partial y^{2}}\right) 
\;+\; i k y \left(\frac{\partial}{\partial x} + i \frac{\partial}{\partial y}\right),
\end{equation}
where $z = x + i y \in \mathbb{H}$. This operator plays a central role in the theory of 
Maass forms, harmonic Maass forms, and their generalizations.

  The requirement \(\Delta_k^{\,d+1} f=0\) provides PMFs with a controlled non-holomorphic dependence (for example through incomplete Gamma functions or powers of \(1/y\)) while preserving modular transformation laws.

Polyharmonic Maass forms (PMFs), as well as harmonic Maass forms, admit Fourier expansions that naturally separate into holomorphic and non-holomorphic contributions. 
A typical decomposition can be expressed as  
\begin{equation}
Y(\tau) \;=\; \sum_{n \in \tfrac{1}{N}\mathbb{Z}_{\geq 0}} c^{+}_{n}\, q^{n} 
\;+\; c^{-}_{0}\, y^{1-k} 
\;+\; \sum_{n \in \tfrac{1}{N}\mathbb{Z}_{<0}} c^{-}_{n}\, 
\Gamma(1-k,-4\pi n y)\, q^{n},
\qquad q \equiv e^{2\pi i \tau},
\end{equation}
where the coefficients $c^{+}_{n}$ correspond to the holomorphic part, while the terms 
involving incomplete gamma functions $\Gamma(1-k,-4\pi n y)$ encode the non-holomorphic 
contributions.

Matter fields are assigned both representations of the finite modular group and modular weights, which collectively determine their transformation properties under $\Gamma_{N}$. Denoting a left-handed chiral matter field by $\psi$ and its modular conjugate by $\psi^{c}$, their transformations under $\gamma \in SL(2,\mathbb{Z})$ 
are given by
\begin{align}
\psi(x) &\;\longrightarrow\; (c\tau + d)^{-k_\psi}\, \rho_{\psi}(\gamma)\, \psi(x), \\
\psi^{c}(x) &\;\longrightarrow\; (c\tau + d)^{-k_{\psi^{c}}}\, \rho_{\psi^{c}}(\gamma)\, \psi^{c}(x),
\end{align}
where $k_\psi$ and $k_{\psi^{c}}$ denote the respective modular weights, and 
$\rho_{\psi}(\gamma)$ and $\rho_{\psi^{c}}(\gamma)$ are the representation matrices 
of $\gamma$.

\begin{table}[ht]
\centering
\begin{tabular}{ll}
\toprule
Weight \(k\) & Available PMF multiplets at level \(N=3\) \\
\midrule
\(k=-4\) & \(Y^{(-4)}_{1},\;Y^{(-4)}_{3}\) \\
\(k=-2\) & \(Y^{(-2)}_{1},\;Y^{(-2)}_{3}\) \\
\(k=0\)  & \(Y^{(0)}_{1},\;Y^{(0)}_{3}\) \\
\(k=2\)  & \(Y^{(2)}_{1},\;Y^{(2)}_{3}\) \\
\(k=4\)  & \(Y^{(4)}_{1},\;Y^{(4)}_{3}\) \\
\(k=6\)  & \(Y^{(6)}_{1},\;Y^{(6)}_{3},\;Y^{(6)}_{3'}\) \\
\bottomrule
\end{tabular}
\caption{Representative polyharmonic Maass form multiplets at level \(N=3\) used in model building. The subscript indicates the \(A_4\) representation (singlet or triplet). The list is schematic and indicates the types of PMFs that appear in the modular expansions employed in the text.}
\label{tab:pmf_weights}
\end{table}

A Yukawa coupling $Y(\tau)$ appearing in the Dirac mass term, which connects fermion fields with the Higgs  must transform in a manner that ensures invariance of the full interaction under the modular group. For a generic Dirac-type term
\begin{equation}
\mathcal{L}_{Y} = Y(\tau)\, \psi^{c} \psi H \;\subset\; Y(\tau)\, \overline{\psi} \psi H,
\end{equation}
modular invariance imposes the following constraints on the modular weights and representations:
\begin{equation}
k_Y = k_{\psi^{c}} + k_{\psi} + k_H, 
\qquad
\rho_Y \otimes \rho_{\psi^{c}} \otimes \rho_{\psi} \otimes \rho_H \;\supset\; \mathbf{1},
\end{equation}
where $\mathbf{1}$ denotes the trivial singlet representation. Similarly, for a Majorana mass term of the form $\psi^{c} \psi^{c}$, one requires
\begin{equation}
k_Y = 2 k_{\psi^{c}}, 
\qquad
\rho_Y \otimes \rho_{\psi^{c}} \otimes \rho_{\psi^{c}} \;\supset\; \mathbf{1}.
\end{equation}
These conditions ensure that the Yukawa interactions respect modular invariance, thereby allowing modular forms to consistently act as flavor-dependent coupling 
constants in the theory.

Thus, the allowed modular Yukawa factors $Y(\tau)$ are polyharmonic Maass forms (PMFs) with definite modular weight and $A_4$ transformation properties. These functions act as effective, $\tau$-dependent Yukawa couplings. At level $N=3$, one can construct 
modular multiplets that transform either as the triplet $\mathbf{3}$ or as the singlets $\mathbf{1}, \mathbf{1}', \mathbf{1}''$ of $A_4$. In the present work, we employ PMFs of several weights, for example $k_Y = 0, -2, \dots$, arranged into triplet structures. 
\begin{equation}
Y^{(k)}_{\mathbf{3}}(\tau) = \big(Y^{(k)}_{3,1},\, Y^{(k)}_{3,2},\, Y^{(k)}_{3,3}\big)^{T}.
\end{equation}
A compact summary of the weights considered in our analysis is provided in Table~\ref{tab:pmf_weights}.

\section{Model framework}
\label{sec:model}
We consider a minimal extension of Standard Model (SM) with inclusion of three right-handed neutrinos (RHNs) $N_{Ri}$ with i=1,2,3 invoked by finite modular symmetry group of level $N=3$ i.e, $\Gamma_3 \simeq A_4$. The interesting feature of implementing $A_4$ modular symmetry is to yield a predictive and simplified structure of leptonic mass matrices, thereby light neutrino masses are generated via type-I seesaw in a minimal way. The usual leptons $\ell_L, \ell_R$, SM Higgs $H$ and RHNs $N_{Ri}$ are presented with their symmetry transformation and modular weight assignment in Table~\ref{tab:fields-inverse}.

The modular-invariant Lagrangian relevant for the lepton masses is written as
\begin{align}
-\mathcal{L}_{M_{A_{4}}}= \mathcal{L}_{M_\ell}+\mathcal{L}_{M_{D}}+\mathcal{L}_{M_R} \, ,
\end{align}
where $\mathcal{L}_{M_\ell}$, $\mathcal{L}_{M_{D}}$ and $\mathcal{L}_{M_R}$ denote the Lagrangian terms for charged leptons, Dirac neutrinos and RH Majorana masses, respectively. 

\begin{table}[h!]
\begin{center}
\begin{tabular} {|c|c|c|c|c|c|}\hline
Fields & $SU(3)_C$ & $SU(2)_L$ & $U(1)_Y$ & $A_4$ & Modular weight $k$ \\ 
\hline
$\{\overline{L_{e_L}}, \overline{L_{\mu_L}}, \overline{L_{\tau_L}} \}$ & {\bf 1} & {\bf 2} & $1/2$ & {\bf 3} & 2 \\
\hline
$\{e_R, \mu_R, \tau_R\}$ & {\bf 1} & {\bf 1} & $-1$ & $\{ {\bf 1}, {\bf 1^{\prime\prime}}, {\bf 1^{\prime}} \}$  & $-2$ \\
\hline
$H$ & {\bf 1} & {\bf 2} & $1/2$ & {\bf 1} & 0  \\
\hline 
$\tilde{H}$ & {\bf 1} & {\bf 2} & $-1/2$ & {\bf 1} & 0  \\
\hline 
$N_{1R}, N_{2R}, N_{3R}$ & {\bf 1} & {\bf 1} & $0$ & $\{ {\bf 1}, {\bf 1^{\prime}}, {\bf 1^{\prime\prime}} \}$ & 0\\
\hline
\end{tabular}
\end{center}
\caption{Particle content and transformation properties of fields under $A_4$ modular symmetry with modular weight assignments.}
\label{tab:fields-inverse}
\end{table}

\subsection{Structure of the charged lepton mass matrix}

The charged lepton mass matrix $M_\ell$ arises from the Yukawa interactions between the left-handed lepton doublets and the right-handed charged leptons with the Higgs field. The structure of $M_\ell$ is determined by the modular forms of appropriate weight. In the present framework, the left-handed leptons are assigned to an $A_4$ triplet, while the right-handed charged leptons can be assigned either to singlets or a triplet representation, depending on the specific model. 

In this work, we assign the three left-handed lepton doublets, $L_\alpha$ $(\alpha = e, \mu, \tau)$, to transform as a triplet under the $A_4$ flavor symmetry. Similarly, the right-handed charged leptons $(e_R, \mu_R, \tau_R)$ are considered to transform as different one dimensional $A_4$ representations i.e, $\{ {\bf 1}, {\bf 1^{\prime \prime}}, {\bf 1^{\prime}} \}$. The Standard Model Higgs doublet $H$ is assumed to be an $A_4$ singlet with $0$ modular weight. The modular weights of the left- and right-handed leptons are chosen as $2$ and $-2$, that ensures modular invariance of the Yukawa interactions..

The charged-lepton Yukawa Lagrangian is given by
\begin{equation}
 \mathcal{L}_{M_\ell}=\alpha_\ell Y^{(0)}_{3} \overline{L}_{L} H e_R 
+ \beta_\ell Y^{(0)}_{3} \overline{L}_{L} H \mu_R 
+ \gamma_\ell Y^{(0)}_{3} \overline{L}_{L} H \tau_R + {\rm h.c.},
 \label{CL}
\end{equation}
where $Y^{(0)}_3$ denotes the $A_4$ modular triplet constructed from polyharmonic Maass forms (PHMFs) of weight zero. The need for flavon fields are replaced by these modular forms and contribute to the role of effective Yukawa couplings. Their explicit Fourier expansions are provided in Eqns (\ref{Y03}), (\ref{Y-23}).  

After electroweak symmetry breaking (EWSB), with $\langle H \rangle = v$, the charged-lepton mass matrix takes the form
\begin{align}
M_\ell = v \begin{pmatrix} 
\alpha_\ell Y^{(0)}_{3,1}  &  \beta_\ell Y^{(0)}_{3,2} &  \gamma_\ell Y^{(0)}_{3,3} \\
\alpha_\ell Y^{(0)}_{3,3}  &  \beta_\ell Y^{(0)}_{3,1}  &  \gamma_\ell Y^{(0)}_{3,2} \\
\alpha_\ell Y^{(0)}_{3,2}  &  \beta_\ell Y^{(0)}_{3,3}  &  \gamma_\ell Y^{(0)}_{3,1} 
\end{pmatrix}.
\label{Eq:Me}
\end{align}
Each parameter has a defined role in this charged lepton mass matrix $M_\ell$. The parameter $v$ denotes the vacuum 
expectation value (VEV) of the Standard Model Higgs doublet, which breaks the electroweak symmetry and settles the overall mass scale of the charged leptons. The coefficients $\alpha_\ell$, $\beta_\ell$, and $\gamma_\ell$ are dimensionless Yukawa cofactors connected with the charged lepton sector. They determine the relative weights of the interactions between the left-handed lepton doublets and the 
right-handed charged leptons, thereby affecting the hierarchical pattern of the charged lepton masses. The quantities $Y^{(0)}_i \ (i=1,2,3)$ represent the components of modular forms of weight zero, which transform as a triplet under the modular group of $A_4$ symmetry. These modular forms are functions of the modulus parameter $\tau$, and they encode the flavor structure of the charged lepton mass matrix by acting as an effective Yukawa coupling. Overall, the modular forms $Y^{(0)}_i$ determine the specific texture of $M_\ell$, while the coefficients $\alpha_\ell, \beta_\ell, \gamma_\ell$ provide freedom to fit the observed charged lepton mass matrix. Thus, the charged lepton mass matrix is formed by the interaction of the Higgs VEV, the Yukawa couplings, and the modular symmetry, leading to a predictive framework for charged lepton masses and mixings.

The charged lepton mass matrix is diagonalized via a bi-unitary transformation,
\begin{equation}
M_\ell= V_L M_{\ell}^{\rm diag} U^{\dagger}_R\, ,
\end{equation}
where $M_\ell^{\rm diag}=\text{Diag}(m_e,m_\mu,m_\tau)$ are charged lepton mass eigenvalues and $V_L\equiv U_\ell$ and $V_R$ are corresponding diagonalizing mixing matrices. However, the appropriate combinations of charged lepton mass matrices associated with a single unitary mixing matrix, relevant for the low-energy neutrino observables, are given by
\begin{eqnarray}
&& {\rm Tr}(M_\ell^\dagger M_\ell)=|m_e|^2+|m_\mu|^2+|m_\tau|^2, \nonumber \\
&& {\rm Det}(M_\ell^\dagger M_\ell)=|m_e|^2 |m_\mu|^2 |m_\tau|^2, \nonumber \\
&& \frac{1}{2}\left[{\rm Tr}(M_\ell^\dagger M_\ell)\right]^2-\frac{1}{2}{\rm Tr}\left[(M_\ell^\dagger M_\ell)^2\right] 
= m_e^2 m_\mu^2 + m_\mu^2 m_\tau^2 + m_\tau^2 m_e^2. 
\label{Constraints}
\end{eqnarray}
These constraints are helpful to calculate the unknown pre-factors $\alpha_\ell, \beta_\ell, \gamma_\ell$ in the diagonalization method.  
\subsection{Structure of the Dirac neutrino mass matrix.}
The Dirac neutrino mass matrix $M_D$ encrypts the couplings between the left-handed neutrinos and the right-handed neutrinos and plays a crucial role in the type-I seesaw mechanism. Its structure is assigned by the $A_4$ modular symmetry and the modular forms associated with the modulus $\tau$. For the neutrino sector, we assign the LH lepton doublets $L_\alpha$ as an $A_4$ triplet, while the RH neutrinos $N_{iR}$ transform as $1$, $1^\prime$, and $1^{\prime\prime}$ representations of $A_4$. The Higgs doublet $H$ is an $A_4$ singlet with zero modular weight. The Yukawa couplings for neutrinos arise from $A_4$ triplet modular forms $Y^{(-2)}_3$ of weight $-2$. The relevant interaction Lagrangian for the Dirac mass term for the neutrinos takes the form:
\begin{align}
  \mathcal{L}_{M_D} = \alpha_{D}\,(\overline{L_{L}} Y^{(-2)}_{3})_{1} N_{1R} \tilde{H}
+ \beta_{D}\,(\overline{L_{L}} Y^{(-2)}_{3})_{1^{\prime\prime}} N_{2R}\tilde{H}
+ \gamma_{D}\,(\overline{L_{L}} Y^{(-2)}_{3})_{1^{\prime}} N_{3R}\tilde{H} + {\rm h.c.}
\label{MD}
\end{align}
Here, $\tilde{H}=i \tau_2 H^*$.  After EWSB, the Dirac neutrino mass matrix is obtained as
\begin{align}
M_D = v \begin{pmatrix} 
\alpha_{D} Y_{3,1}^{(-2)}  &  \beta_{D} Y_{3,3}^{(-2)} &  \gamma_D Y_{3,2}^{(-2)} \\
\alpha_{D} Y_{3,3}^{(-2)}  & \beta_{D} Y_{3,2}^{(-2)}  &  \gamma_D Y_{3,1}^{(-2)} \\
\alpha_{D} Y_{3,2}^{(-2)}  & \beta_{D} Y_{3,1}^{(-2)}  &  \gamma_D Y_{3,3}^{(-2)}
\end{pmatrix}.
\label{Eq:Md}
\end{align}
The overall scale of the Dirac mass matrix is determined by the vacuum expectation value (VEV) of the SM Higgs doublet, which breaks the electroweak symmetry and provides mass terms to the fermions. The coefficients $\alpha_D$,$\beta_D$, and $\gamma_D$ represent dimensionless Yukawa pre-factors associated with the Dirac neutrino sector. These couplings serve as free parameters that control the relative strengths of the interactions between the left-handed lepton doublets and the right-handed neutrino fields, thereby influencing the structure of $M_D$. The modular forms $Y^{(-2)}_i(\tau)$ with $i=1,2,3$ transform as a triplet under the $A_4$ modular group and carry modular weight $-2$. They are holomorphic functions of the modulus parameter $\tau$. In general, the structure of $M_D$ is constructed by SM Higgs VEV ($v$), the Yukawa pre-factor parameters $\alpha_D$, $\beta_D$, $\gamma_D$, and the modular forrm Yukawa triplets $Y^{(-2)}_{3,i}(\tau)$. Thus the Higgs VEV sets the mass scale, the modular forms introduce $\tau$-dependent flavor correlations, and the Yukawa couplings provide flexibility to reproduce the observed neutrino masses and mixing patterns.
\subsection{Structure of the Majorana mass matrix for RH neutrinos.}
The heavy right-handed neutrinos obtain a Majorana mass term through the breaking of the underlying modular $A_4$ symmetry, which plays a crucial role in the type-I seesaw mechanism. The structure of the Majorana mass matrix $M_R$ is given by the $A_4$ modular symmetry and the corresponding modular forms, ensuring a predictive appearance for neutrino masses. The Majorana mass term for the RHNs is generated through the $A_4$ singlet modular form $Y^{(0)}_1$. The invariant Lagrangian takes the form
\begin{align}
  \mathcal{L}_{M_R} = \Lambda_1 Y^{(0)}_{1} N_{1R} N_{1R} 
  + \Lambda_2 Y^{(0)}_{1}(N_{2R} N_{3R} + N_{3R} N_{2R}) + {\rm h.c.}
  \label{MR}
\end{align}
The resulting structure of the RHN Majorana mass matrix, by fixing $Y^{(0)}_{1} \simeq 1$, takes the form
\begin{equation}
M_R = \begin{pmatrix}
 \Lambda _1 & 0 & 0 \\
 0 & 0 & \Lambda _2 \\
 0 & \Lambda _2 & 0 
\end{pmatrix}.
\label{Eq:Mn} 
\end{equation}
The right-handed neutrino Majorana mass matrix $M_R$ has a constrained structure imposed by the $A_4$ flavor symmetry. The parameter $\Lambda_1$ denotes the bare (does not require the Higgs mechanism or any scalar field)  Majorana mass of the $A_4$ singlet right-handed neutrino. At the same time, $\Lambda_2$ defines the masses of the two non-trivial $A_4$ representations. Thus, only two independent parameters, $\Lambda_1$ and $\Lambda_2$, control the Majorana sector, leading to a predictive structure that plays a crucial role in the 
seesaw mechanism.
\subsection{Light neutrino masses via type-I seesaw}
Using derived structure of Dirac neutrino mass matrix $M_D$ and RHN mass matrix ($M_R$), the resulting light neutrino mass matrix is obtained by applying the type-I seesaw formula:
\begin{eqnarray}
m_\nu &=& -M_D M_R^{-1} M_D^T \nonumber 
\label{Eq:Mnu1}
\end{eqnarray}
The explicit entries $m_{\nu_{ij}} \equiv m_\nu[i,j]$ can be expressed in terms of the model parameters $(\alpha_D, \beta_D, \gamma_D, \Lambda_1, \Lambda_2)$ and modular forms $Y^{(-2)}_{3,i}$. These relations form the basis for the numerical analysis, where we will fit neutrino oscillation data and extract phenomenological predictions in next section. 

\begin{equation}
m_\nu \;=\; -\,M_D\,M_R^{-1}\,M_D^T \;=\;
\begin{pmatrix}
m_\nu[1,1] & m_\nu[1,2] & m_\nu[1,3] \\
m_\nu[2,1] & m_\nu[2,2] & m_\nu[2,3] \\
m_\nu[3,1] & m_\nu[3,2] & m_\nu[3,3]
\end{pmatrix},
\label{Eqn:mnu}
\end{equation}
with $m_\nu[i,j]=m_\nu[j,i]$. For simplicity, we denote the three components of the modular triplet by
$$
Y_1 \equiv Y^{(-2)}_{3,1}(\tau),\qquad
Y_2 \equiv Y^{(-2)}_{3,2}(\tau),\qquad
Y_3 \equiv Y^{(-2)}_{3,3}(\tau)\,.
$$
the individual matrix elements are found to be,
\begin{align}
m_\nu[1,1] & = -v^2\!\left(\frac{\alpha_D^2}{\Lambda_1}\,Y_1^2 \;+\; \frac{2\,\beta_D\gamma_D}{\Lambda_2}\,Y_2 Y_3\right), \nonumber \\
m_\nu[2,2] & = -v^2\!\left(\frac{\alpha_D^2}{\Lambda_1}\,Y_3^2 \;+\; \frac{2\,\beta_D\gamma_D}{\Lambda_2}\,Y_1 Y_2\right),  \nonumber \\
m_\nu[3,3] & = -v^2\!\left(\frac{\alpha_D^2}{\Lambda_1}\,Y_2^2 \;+\; \frac{2\,\beta_D\gamma_D}{\Lambda_2}\,Y_1 Y_3\right),
\end{align}
and the off-diagonal elements (symmetric) are
\begin{align}
m_\nu[1,2] &= -v^2\!\left(\frac{\alpha_D^2}{\Lambda_1}\,Y_1 Y_3 \;+\; \frac{\beta_D\gamma_D}{\Lambda_2}\, \big(Y_2^2 + Y_1 Y_3\big)\right),
\nonumber \\[4pt]
m_\nu[1,3] &= -v^2\!\left(\frac{\alpha_D^2}{\Lambda_1}\,Y_1 Y_2 \;+\; \frac{\beta_D\gamma_D}{\Lambda_2}\, \big(Y_1 Y_2 + Y_3^2\big)\right),
\nonumber \\[4pt]
m_\nu[2,3] &= -v^2\!\left(\frac{\alpha_D^2}{\Lambda_1}\,Y_2 Y_3 \;+\; \frac{\beta_D\gamma_D}{\Lambda_2}\, \big(Y_1^2 + Y_2 Y_3\big)\right).
\end{align}
It is crucial to highlight that the flavor structure is responsible for generating the light neutrino mass matrix via the type-I seesaw mechanism which also control the interactions of the heavy RHNs. In particular the Dirac Yukawa couplings encoded in $M_D$ and the mass structure of $M_R$ decide both the magnitude of CP violation and the strength of washout effects which is suitable for the thermal leptogenesis. The decay of the lightest RHN creates a lepton asymmetry that is subsequently converted into the observed baryon asymmetry of the Universe through non-perturbative sphaleron processes. Within this framework, the modular $A_4$ symmetry provides the predictive structure for neutrino masses and mixing consistent with oscillation data and also setup direct connection between low-energy neutrino parameters and the high-scale dynamics underlying the matter–antimatter asymmetry.In the subsequent sections, we illustrate that this construction numerically recognize successful leptogenesis while remaining consistent with current neutrino data.

\begin{table}[htb!]
\centering
\begin{tabular}{|c|c|c|}
\hline
\multicolumn{3}{|c|}{Standard oscillation parameters (NuFIT 6.0)} \\
\hline\hline
Parameter & Best-fit value & $3\sigma$ range  \\
\hline\hline
$\theta_{12}({}^{\circ})$  & $33.68^\circ$ & $31.63^\circ \rightarrow 35.95^\circ$  \\
\hline
$\theta_{13}({}^{\circ})$  & $8.52^\circ$  (NO) & $8.18^\circ \rightarrow 8.87^\circ$ (NO)   \\
 & $8.58^\circ$ (IO) & $8.24^\circ \rightarrow 8.91^\circ$ (IO)\\
 \hline
$\theta_{23}({}^{\circ})$ & $48.5^\circ$ (NO) & $41.0^\circ \rightarrow 50.5^\circ$  (NO) \\
 & $48.6^\circ$  (IO) & $41.4^\circ \rightarrow 50.6^\circ$ (IO)
 \\
 \hline
$\Delta m_{21}^2~(\text{eV}^2)$ & $7.49 \times 10^{-5}$ & $(6.92 \rightarrow 8.05)\times 10^{-5}$ \\
\hline
$\Delta m_{3l}^2~(\text{eV}^2)$ & $2.534 \times 10^{-3}$ (NO) & $(2.45 \rightarrow 2.57)\times 10^{-3}$ (NO) \\
 & $-2.484 \times 10^{-3}$ (IO) & $(-2.55 \rightarrow -2.42)\times 10^{-3}$ (IO) \\
 \hline
 $\delta({}^{\circ})$ & $177^{\circ}$  (NO) & $96^{\circ} - 422^{\circ}$  (NO)\\  
  &  $285^{\circ}$  (IO) & $201^{\circ} - 348^{\circ}$  (IO) \\
\hline
\end{tabular}
\caption{Global-fit values of neutrino oscillation parameters (NuFIT 6.0)~\cite{Esteban:2024eli}. Here we use the notation $\Delta m^2_{3l}$ with $l=1$ for  $\Delta m^2_{3l}>0$ in case of normal ordering and $l=2$ for  $\Delta m^2_{3l}<0$ in case of inverted ordering.}
\label{oscparams}
\end{table}
\section{Numerical Analysis of Neutrino Masses and Mixings}
\label{sec:numerical}
In this section, we will perform a systematic numerical study of the neutrino masses and mixing within the modular $A_4$  type-I seesaw framework. The numerical analysis is based on scanning all the input model parameters randomly and derive the allowed parameter sets satisfying all the neutrino oscillation data.  The main goal of the numerical study is divided into three steps: (i) to scan the input model parameters, including the modular parameter $\tau$, within their allowed range that can reproduce oscillation data, (ii) perform a $\chi^2$ fitting and extract correlations among model parameters. The idea is to examine the implications of $A_4$ modular symmetry and the complex modular parameter $\tau$  on the neutrino masses and mixing, Dirac CP phase, and their implications for high-scale thermal leptogenesis.  

We chose input model parameters and complex modulus parameter $\tau$ in the following ranges:
\begin{align}
&|{\rm Re}[\tau]| \in [0, 0.5],\quad {\rm Im}[\tau]\in [0.9, 1.2] \nonumber
\end{align}
As discussed previously in section \ref{sec:model}, we assign non-holomorphic modular weights to the fields as shown in Table \ref{tab:fields-inverse}. We have used Yukawa couplings of weight $k=0,-2$ in singlet and triplet representations to build our desired Lagrangians of charged and neutral lepton sectors as presented in Eqns. (\ref{CL}), (\ref{MD}) and (\ref{MR}) respectively. We have used type-I seesaw model to generate the light neutrino mass matrix for our model. 

To begin the numerical analysis, we chose the values of ${\rm Re}[\tau]$ and ${\rm Im}[\tau]$ over a range of $|{\rm Re}[\tau]| \in$[0, 0.5] and ${\rm Im}[\tau] \in$ [0.9, 1.2] such that it satisfies the constraints of the most generic fundamental domain i.e, $|{\rm Re}[\tau]|\leq 1/2$, ${\rm Im}[\tau] > 0$ and $|\tau| \geq 1$. The Yukawa couplings used in our model are evaluated using the values of ${\rm Re}[\tau]$ and Im($\tau$) from their q-expansions as given in Eqns (\ref{Y03}), (\ref{Y-23}).  We also fix phenomenologically motivated other model parameters like $|\alpha_D,| \; |\beta_D|, \; |\gamma_D|$ over a range of [0.1, 0.99] and $\Lambda_1, \; \Lambda_2$ for a range of $\Lambda_1 \in [1.0 \times 10^{13}, 9\times 10^{13}]$~GeV and $\Lambda_2 \; \in \; [1.0 \times 10^{14}, 9\times 10^{14} ]$~GeV, respectively. These input values are then used to find the numerical values of $M_D$, $M_R$ and eigenvalues of the neutrino mass matrix (Eqn. \ref{Eq:Mnu1}). 

After calculating the eigenvalues of $m_\nu$, we  choose only those eigenvalues of $m_\nu$ i.e, $m_1, \; m_2,\; m_3$ which satisfy two known mass square difference parameters, $\Delta m^2_{21}$ and $\Delta m^2_{31}$ lying within $3\sigma$ range of oscillation parameters presented in Table~\ref{oscparams}. The light neutrino Majorana mass matrix can be diagonalized with the help of a unitary mixing matrix, 
\begin{equation}
    m^{\rm diag}_\nu= U^T_{ \nu} m_\nu U_{ \nu} 
= \mbox{diag}\left(m_1,m_2,m_3 \right)\,,~~m_i > 0
\label{Mnu:diag}
\end{equation}
As can be seen from Eqn (\ref{Eqn:mnu}) that $m_{\nu}$ is a symmetric matrix, finding its normalized eigenvectors numerically gives us $U_{\nu}$ under the transformation Eqn (\ref{Mnu:diag}). It is known that the Pontecorvo, Maki, Nakagawa, Sakata (PMNS) mixing matrix $U_{\rm PMNS}$~ 
\cite{Pontecorvo:1957qd,Maki:1962mu,Pontecorvo:1967fh} is related to the diagonalizing matrices $U_{\nu} $(diagonalizing $m_\nu$)and $U_\ell$ (can be obtained from Eqn.(\ref{Constraints})) as
\begin{equation}
    U_{\rm PMNS}= U^{\dagger}_{\ell}\, U_{\nu}
    \label{PMNS}
\end{equation}
The standard parametrization of the PMNS~\cite{ParticleDataGroup:2018ovx} mixing matrix is given by
\begin{eqnarray}
&&U_{\rm {PMNS}}= 
\begin{pmatrix} c_{13}c_{12}&c_{13}s_{12}&s_{13}e^{-i\delta}\\
-c_{23}s_{12}-c_{12}s_{13}s_{23}e^{i\delta}&c_{12}c_{23}-s_{12}s_{13}s_{23}e^{i\delta}&s_{23}c_{13}\\
s_{12}s_{23}-c_{12}c_{23}s_{13}e^{i\delta}&-c_{12}s_{23}-s_{12}s_{13}c_{23}e^{i\delta}&c_{13}c_{23}
\end{pmatrix} \mbox{P}
\label{Eq:UPMNS} 
\end{eqnarray}
where $c_{ij} = \cos{\theta_{ij}}, \; s_{ij} = \sin{\theta_{ij}}$ are cosine and sine angle of different neutrino mixing angles and $\delta$ is the Dirac CP phase. Additionally, the diagonal phase matrix $\mbox{P}=\mbox{diag}\left( e^{i\alpha}, e^{i \beta}, 1 \right)$ contains the two Majorana CP violating phases $\alpha$ and $\beta$ \cite{Bilenky:1980cx}.


We incorporate the values of the neutrino mixing angles, $\theta_{12},\; \theta_{13}$, $\theta_{23}$ from NuFIT 6.0 and span the values of the Dirac CP phase $\delta_{CP}$ and  Majorana phases $\alpha$ and $\beta$ from $\alpha, \; \beta, \; \delta_{CP} \;\in \; [0,2\pi]$. Thus we get the values of the $U_\ell$ elements from this numerical analysis by using Eqn. (\ref{Constraints})
\begin{equation}
    U^{\dagger}_\ell M_\ell M^{\dagger}_\ell U_\ell = \mbox{diag} (m^2_e ,\; m^2_{\mu}, \; m^2_{\tau})
    \label{mdiag:sq}
\end{equation}
We perform a parameter scan of $m_e, \; m_{\mu}, \; m_{\tau}$ within the limits as given in PDG 2024~\cite{ParticleDataGroup:2024cfk} and solve the set of simultaneous equations obtained from Eqn. (\ref{Constraints}) for obtaining $U_\ell$. The solutions to this set of equations give us the value of the coefficients $ \alpha_l $, $ \beta_l $, $ \gamma_l $.

\begin{table}[t!]
\begin{center}
\begin{tabular}{|c|c|c|} 
\hline 
Model Parameters & \pmb{BP1}  & \pmb{BP2}  \\ \cline{2-3}
\hline \hline
    $m_1$ (eV) & 0.001  & 0.0014 \\ 
    $m_2$  (eV)  &  0.009 & 0.0089\\ 
    $m_3$  (eV)  &  0.05  & 0.049  \\ \cline{2-3}
\hline
$|\alpha_D|$   & 0.1077     & 0.5227  \\ 
$|\beta_D|$   &  0.1229     & 0.4362  \\ 
$|\gamma_D|$   &  0.0733    & 0.5898 \\ \cline{2-3}
\hline
    $Y^{(-2)}_{3,1}$    &  0.42 - 0.015 i   &  0.41 - 0.014 i  \\ 
    $Y^{(-2)}_{3,2}$  & -0.095 + 0.094 i    &  -0.098+ 0.089 i \\ 
    $Y^{(-2)}_{3,3}$  &  -0.255 + 0.1313 i  &  -0.255 + 0.1250 i  \\ \cline{2-3}
    \hline
    $\Lambda_1$ (GeV) & $3.70\times10^{12}$    &  $ 8.688\times10^{13}$ \\ 
    $\Lambda_2$ (GeV) &  $2.354\times 10^{13}$ &  $6.673\times 10^{14}$ \\ \cline{2-3}
    \hline
$Re(\tau) \; ;\; |Re(\tau)| \leq \dfrac{1}{2}$  &  0.219   & 0.21  \\ 
$\Im(\tau)  \; ;\; Im(\tau) >0$  & 0.996                   & 0.99   \\ \cline{2-3}
\hline
$\Delta^2 m_{21}$  & $7.97\times 10^{-5}$     &  $7.82\times 10^{-5}$    \\ 
$\Delta^2 m_{31}$  & $2.54 \times 10^{-3}$    &  $2.48 \times 10^{-3}$   \\ \cline{2-3}
\hline
\end{tabular}
\caption{Representative benchmark points of parameters obtained from our model  with different choice of derived $M_D$ and fixed value of $M_R$. The parameters used for numerical analysis includes BP1 (2nd column) with  Low $M_D$, $\Lambda_1 \approx \mathcal{O}(10^{12}\; \mbox{GeV}) \; \text{and} \;\; \Lambda_2 \approx \mathcal{O}(10^{13}\; \mbox{GeV})$ and BP2 (3rd column) with High $M_D$, $\Lambda_1 \approx \mathcal{O}(10^{13}\; \mbox{GeV}) \; \text{and} \;\; \Lambda_2 \approx \mathcal{O}(10^{14}\; \mbox{GeV})$.}
\label{BPs}
\end{center}
\end{table}
\subsection{Neutrino Observables and benchmark points}
We summarize here the numerical strategy adopted to explore both low and high $M_D$ benchmark regimes as follows:
\begin{itemize}
\item  Compute $Y_i(\tau)$ using chosen range of modulus parameter $\tau$ and estimate $M_D,M_R$ via Eqns.~\eqref{Eq:Md}--\eqref{Eq:Mn}.
  \item Construct structure of light neutrino mass $m_\nu$ via type-I seesaw mechanism as given in Eq.(\ref{Eqn:mnu}) and diagonalize to obtain the mass eigenvalues $m_i$ and mixing matrix $U_\nu$. 
  \item Derive numerical structure of $U_\ell$ by diagonalizing $M_\ell M_\ell^\dagger $ (charged-lepton sector) and form $U_{\rm PMNS}$ using Eq.~(\eqref{Eq:UPMNS}).
  \item Extract the oscillation observables like mixing angles $\{\theta_{12},\theta_{13},\theta_{23}\}$, two mass square difference parameters $\Delta m^2_{21}, \Delta m^2_{3\ell}$ and Dirac CP phase $\delta$. 
\end{itemize}

\begin{figure}[htb!]
  \centering
    \includegraphics[width=\linewidth]{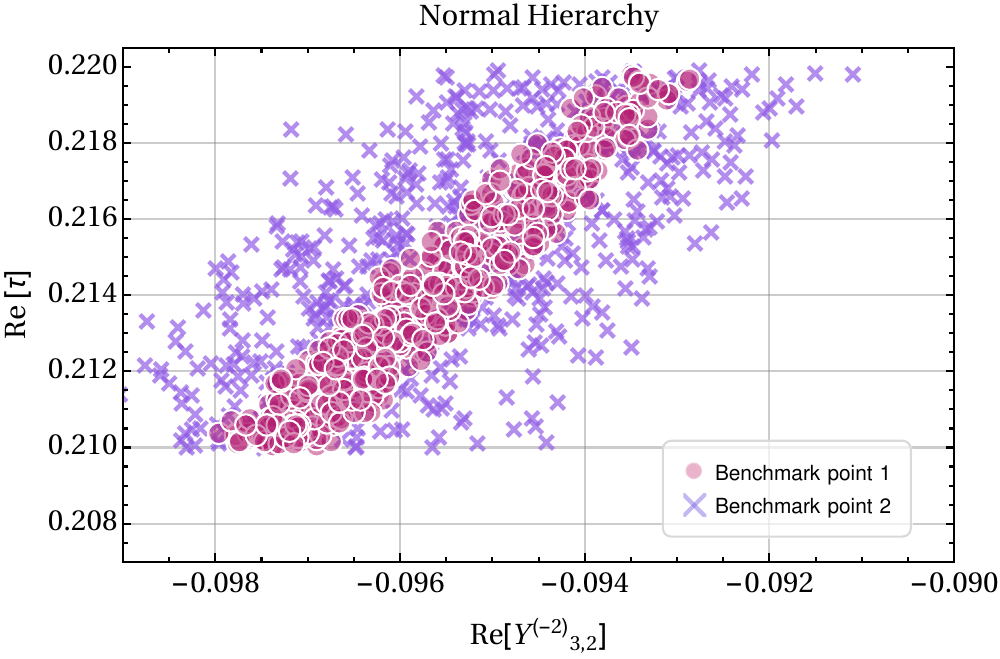}
  \caption{Plots showing the correlation among  Re[$Y^{(-2)}_{3,2}$] vs Re[$\tau$] that exist in the upper half plane, $\mathcal{H}$. The \enquote{$\textcolor{RedViolet}{\bullet}$} and \enquote{ $\textcolor{Orchid}{\bm{\times}}$} color coded points  correspond to the  benchmark points BP1 ($ M_D \approx \; \mathcal{O}(1\; GeV) $) and BP2 ($ M_D \approx \; \mathcal{O}(10\; GeV) $) respectively as presented in Table.\ref{tab:BP1} and Table.\ref{tab:BP2}.}
  \label{fig1}
\end{figure}
\begin{table}[t]
    \centering
    \resizebox{\linewidth}{!}{
    \begin{tabular}{c c c c c }
       \hline
     \rowcolor{blue!10}   \textcolor{blue}{$m_1$ (eV)} &  \textcolor{blue}{$m_2$ (eV)} &  \textcolor{blue}{$m_3$ (eV) } &  \textcolor{blue}{$\Delta m^2_{21}/10^{-5} (\text{eV}^2)$} &  \textcolor{blue}{$\Delta m^2_{31}/10^{-3} (\text{eV}^2)$ }\\
        \midrule
    $\{0.0014,0.0016\}$   & $\{0.0087,0.0091\}$ &  $ \{0.0496,0.0509\}$ & $\{7.37,8.05\}$ & $\{2.463,\; 2.590\}$  \\
        \midrule
   \rowcolor{blue!10}      \textcolor{blue}{ $\delta_{CP} (^\circ)$ }  & \textcolor{blue}{$Re(\tau) $} & \textcolor{blue}{ $Im (\tau)$ }  &  \textcolor{blue}{$\alpha (^\circ) $ }  & \textcolor{blue}{ $\beta (^\circ)$ } \\
          \midrule
         $\{0.00002,6.28\}$  & $\{0.21,0.219\}$  & $\{0.995,0.997\}$  &  $\{0.0037,\; 360 \}$ &  $\{0.000049,\; 360\}$ \\
         \midrule
  \rowcolor{blue!10}  \textcolor{blue}{$\Lambda_1/10^{21}$  (GeV)} &  \textcolor{blue}{$\Lambda_2/10^{22}$ (GeV)} &  \textcolor{blue}{   $|\alpha_D|$} &  \textcolor{blue}{$|\beta_D|$ } &  \textcolor{blue}{$|\gamma_D|$}     \\
          \midrule
   $\{3.7 ,3.7002\}$  & $\{2.353,2.355\}$ & $\{0.003,0.5\}$ & $\{0.006,0.609\}$  &  $\{0.003,0.553\}$    \\
         \midrule
   \rowcolor{blue!10}     \textcolor{blue}{ $ |\alpha_l/\beta_l| $} &   \textcolor{blue}{$ |\beta_l/\gamma_l| $} &   & & \\
        \midrule
    $\{0.866,0.880\}$  &  $\{1.65,1.70\}$ &  &   & \\
      \bottomrule
    \end{tabular}}
    \caption{Ranges of the various model parameters for $M_D \sim \mathcal{O} ( 1 \; GeV) $ (i.e., benchmark point 1) as deduced from our Type-I $A_4$ model induced with modular symmetry.}
    \label{tab:BP1}
\end{table}
\begin{table}[t]
    \centering
     \resizebox{\linewidth}{!}{
    \begin{tabular}{c c c c c }
       \hline
     \rowcolor{blue!10}   \textcolor{blue}{$m_1$ (eV)} &  \textcolor{blue}{$m_2$ (eV)} &  \textcolor{blue}{$m_3$ (eV) } &  \textcolor{blue}{$\Delta m^2_{21}/10^{-5} (\text{eV}^2)$} &  \textcolor{blue}{$\Delta m^2_{31}/10^{-3} (\text{eV}^2)$ }\\
        \midrule
    $\{0.0014,0.0017 \}$   & $\{0.008, 0.009\}$ &  $ \{0.049,0.0510 \}$ & $\{6.99,8.05\}$ & $\{2.463,\; 2.606\}$  \\
        \midrule
   \rowcolor{blue!10}      \textcolor{blue}{ $\delta_{CP} (^\circ)$ }  & \textcolor{blue}{$Re(\tau) $} & \textcolor{blue}{ $Im (\tau)$ }  &  \textcolor{blue}{$\alpha (^\circ) $ }  & \textcolor{blue}{ $\beta (^\circ) $ } \\
          \midrule
         $\{96.0, 422.0 \}$  & $\{0.21,0.22\}$  & $\{0.99, 0.999\}$  &  $\{0.0037,360 \}$ &  $\{0.00005,360\}$ \\
         \midrule
  \rowcolor{blue!10}  \textcolor{blue}{$\Lambda_1/10^{22}$  (GeV)} &  \textcolor{blue}{$\Lambda_2/10^{23}$ (GeV)} &  \textcolor{blue}{   $|\alpha_D|$} &  \textcolor{blue}{$|\beta_D|$ } &  \textcolor{blue}{$|\gamma_D|$}     \\
          \midrule
   $\{8.68 ,\; 8.7\}$  & $\{6.6,\; 6.67 \}$ & $\{0.006, 0.534\}$ & $\{0.005, 0.778\}$  &  $\{0.005, 0.669\}$    \\
         \midrule
   \rowcolor{blue!10}     \textcolor{blue}{ $ |\alpha_l/\beta_l| $} &   \textcolor{blue}{$ |\beta_l/\gamma_l| $} &   & & \\
        \midrule
    $\{1.19,\; 1.23\}$  &  $\{0.72,\; 0.75\}$ &  &   & \\
      \bottomrule
    \end{tabular}}
    \caption{Ranges of the various model parameters for $M_D \sim \mathcal{O} ( 10 \; GeV) $ (i.e., benchmark point 2) as deduced from our Type-I $A_4$ model induced with modular symmetry.}
    \label{tab:BP2}
\end{table}

\begin{figure}[htb!]
  \centering
    \includegraphics[width=\linewidth]{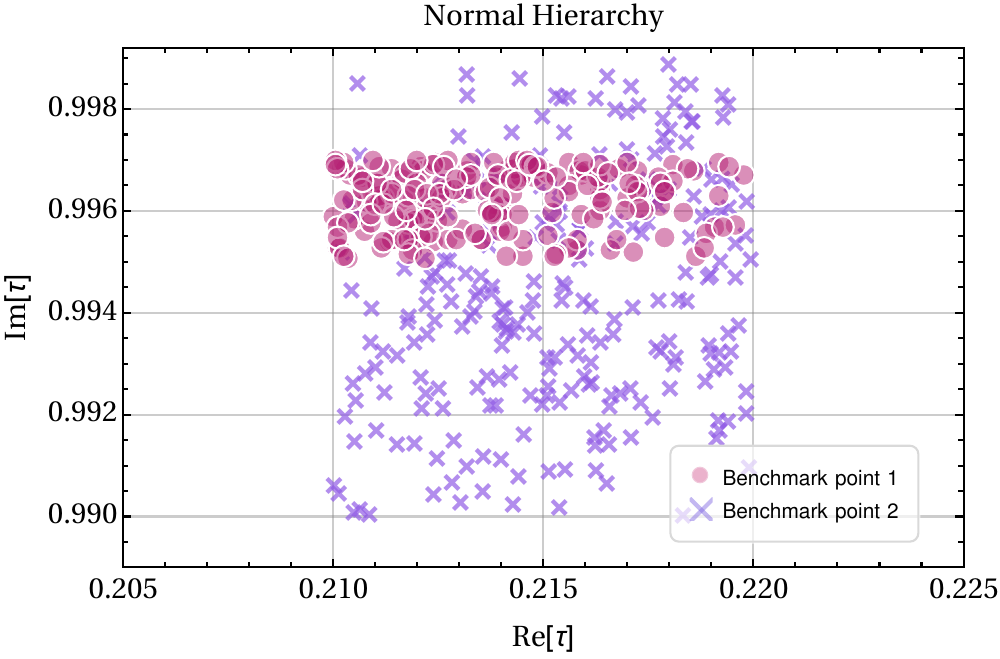}
  \caption{ Plots showing the correlation between Re[$\tau$] vs Im[$\tau$].  The \enquote{$\textcolor{RedViolet}{\bullet}$} and \enquote{ $\textcolor{Orchid}{\bm{\times}}$} color coded points  correspond to the  benchmark points BP1 and BP2 respectively as presented in Table.  \ref{tab:BP1} and Table. \ref{tab:BP2}.}
  \label{fig2}
\end{figure}

Below we present two benchmark points of the Dirac mass matrix where in the first case we take $\Lambda_1 \approx O(10^{12})\; GeV$ and  $\Lambda_2 \approx O(10^{13})\; GeV$. The corresponding values of the other parameters are shown in Table \ref{BPs}. Similarly, in the second case we present $\Lambda_1 \approx O(10^{13})\; GeV$ and  $\Lambda_2 \approx O(10^{14})\; GeV$and its corresponding parameter values are shown in Table \ref{BPs}. We shall use these benchmark points to evaluate the CP asymmetry in the leptogenesis section.
\begin{eqnarray}
\textbf{BP1:}\quad &&M_D = \left(      \begin{matrix}
     -11.14 - i \; 0.405  &  -7.723+ i\; 3.973 &  -1.729 + i\; 1.709 \\
 -6.768 + i \; 3.482      &  -2.896 +i\; 3.482 & 7.591 -i\; 0.276 \\
      -2.538 +i\; 2.509 &  12.71 -i\; 0.463 & -4.610 + i\; 2.372 \\
    \end{matrix}\right) \;\mbox{GeV} 
\nonumber \\
  &&M_R = \left(      \begin{matrix}
     3.70\times10^{12} &   0 &  0 \\
 0     & 0 & 2.354\times 10^{13} \\
   0 &  2.354\times 10^{13}  &0 \\
    \end{matrix}\right) \;\mbox{GeV} 
\end{eqnarray}

\begin{eqnarray}
\textbf{BP2:}\quad &&M_D = \left(      \begin{matrix}
     53.29 - i \; 1.89  &  -27.45+ i\;13.41 &  -14.32 + i\; 13.05 \\
 -32.89 + i \; 16.07      &  -10.59 +i\; 9.65 & 60.12 -i\; 2.13 \\
      -12.69 +i\; 11.57 &  44.47 -i\; 1.57 & -37.12 + i\; 18.14 \\
    \end{matrix}\right) \;\mbox{GeV} 
\nonumber \\
  &&
  M_R = \left(      \begin{matrix}
     8.688\times10^{13} &   0 &  0 \\
 0     & 0 & 6.673\times 10^{14} \\
   0 &   6.673\times 10^{14} &0 \\
    \end{matrix}\right) \;\mbox{GeV} 
\end{eqnarray}
\begin{figure}[h!]
  \centering
    \includegraphics[width=\linewidth]{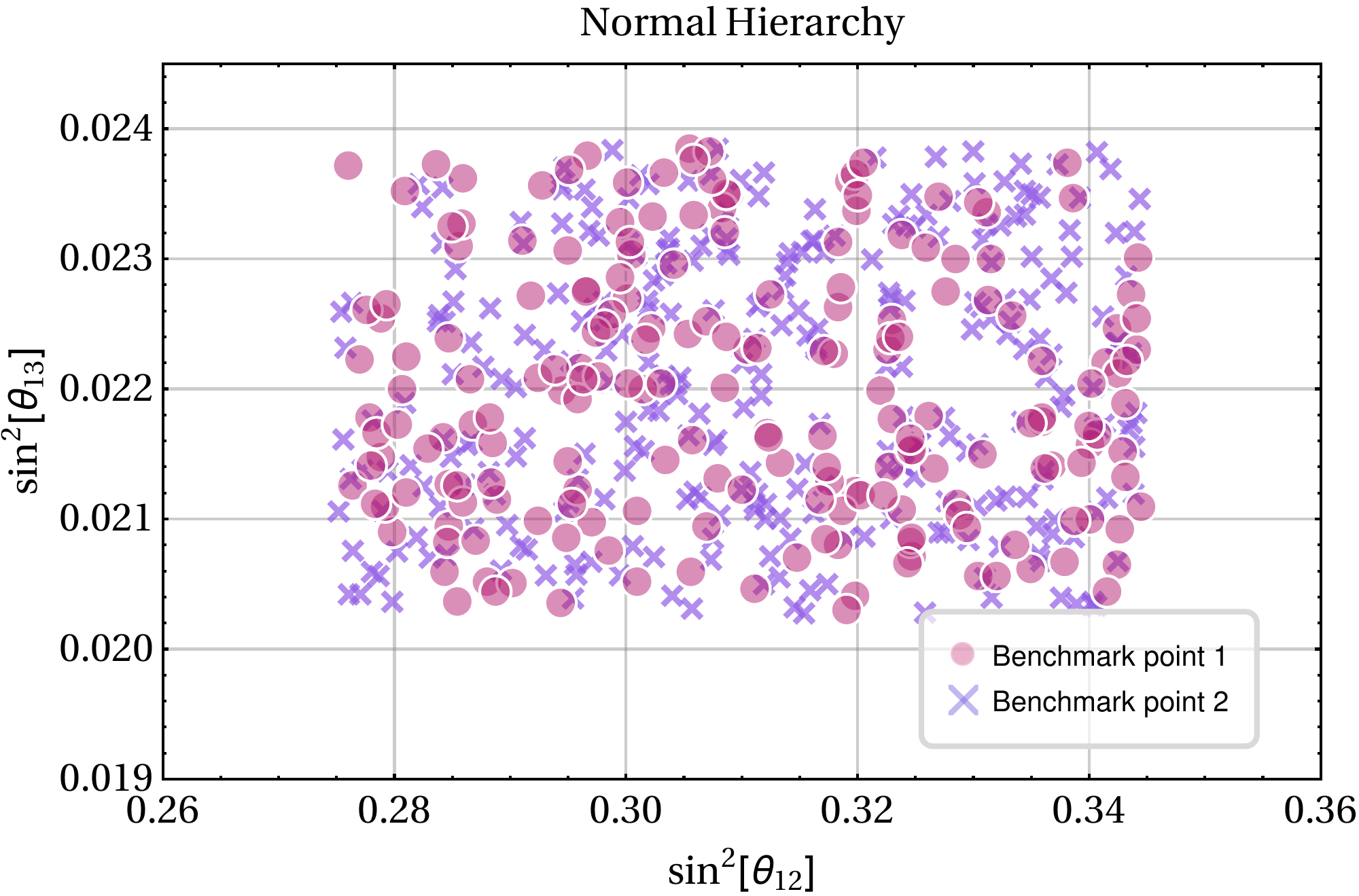}
  \caption{Correlation plots between $\sin^2\theta_{12}$ and $\sin^2\theta_{13}$ are shown here where the measured values of $\theta_{12}$ and $\theta_{13} $ are taken from the NuFIT6.0 for $3\sigma$ range. The \enquote{$\textcolor{RedViolet}{\bullet}$} and \enquote{ $\textcolor{Orchid}{\bm{\times}}$} color coded points  correspond to the  benchmark points BP1 and BP2 respectively as presented in Table. \ref{tab:BP1} and Table. \ref{tab:BP2}.}
  \label{fig3}
\end{figure}
\subsection{Correlation plots and modular dependence}

\begin{figure}[htb!]
  \centering
    \includegraphics[width=\linewidth]{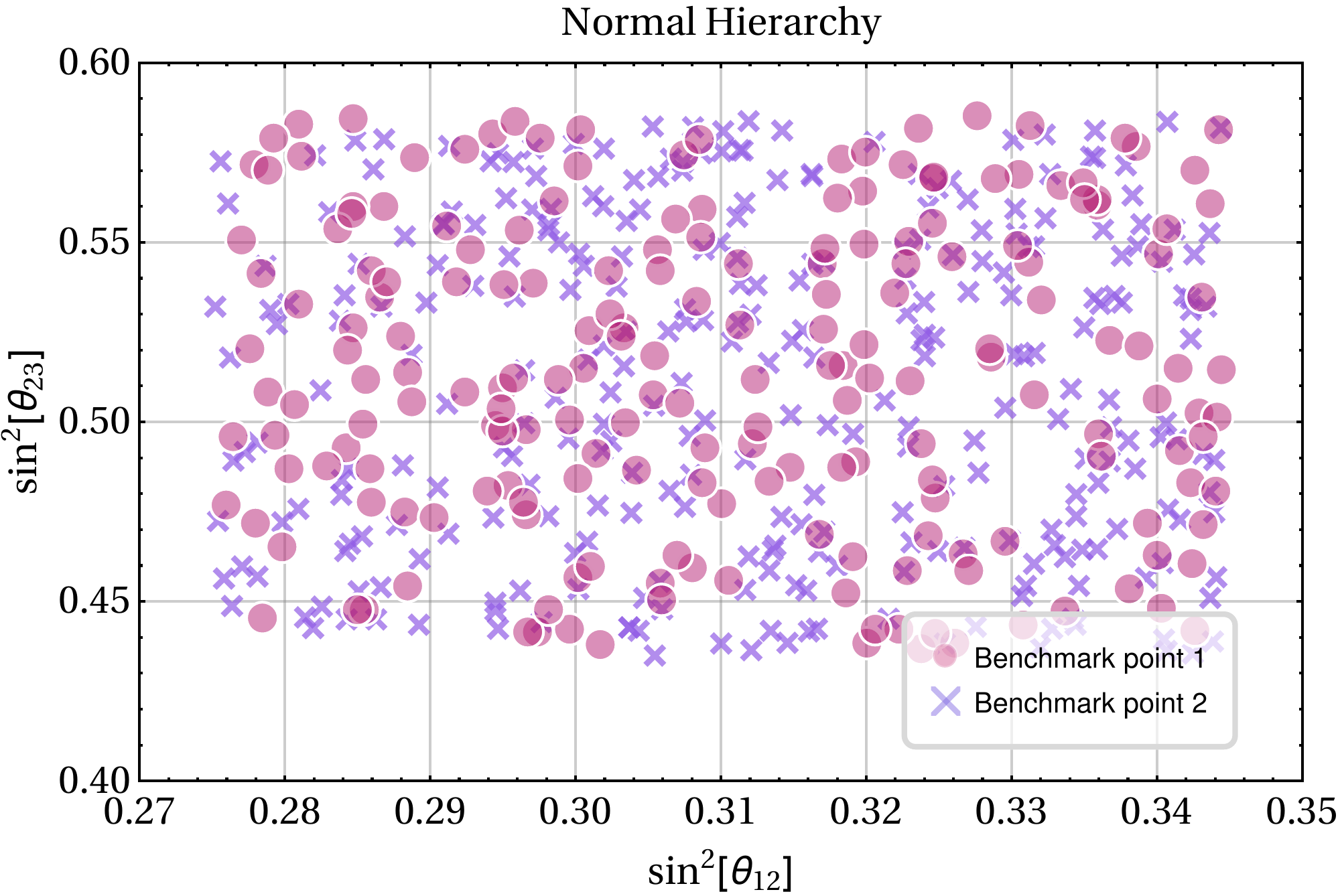}
  \caption{Plot showing correlations between  between $\sin^2\theta_{12}$ and $\sin^2\theta_{23}$. We have taken the $3\sigma $ values of $\theta_{12}$ and $\theta_{23}$ from NuFIT6.0 (2024). The \enquote{$\textcolor{RedViolet}{\bullet}$} and \enquote{ $\textcolor{Orchid}{\bm{\times}}$} color coded points  correspond to the  benchmark points BP1  and BP2  respectively as presented in Table. \ref{tab:BP1} and Table. \ref{tab:BP2}.}
  \label{fig4}
\end{figure}

\begin{figure}[htb!]
  \centering
    \includegraphics[width=\linewidth]{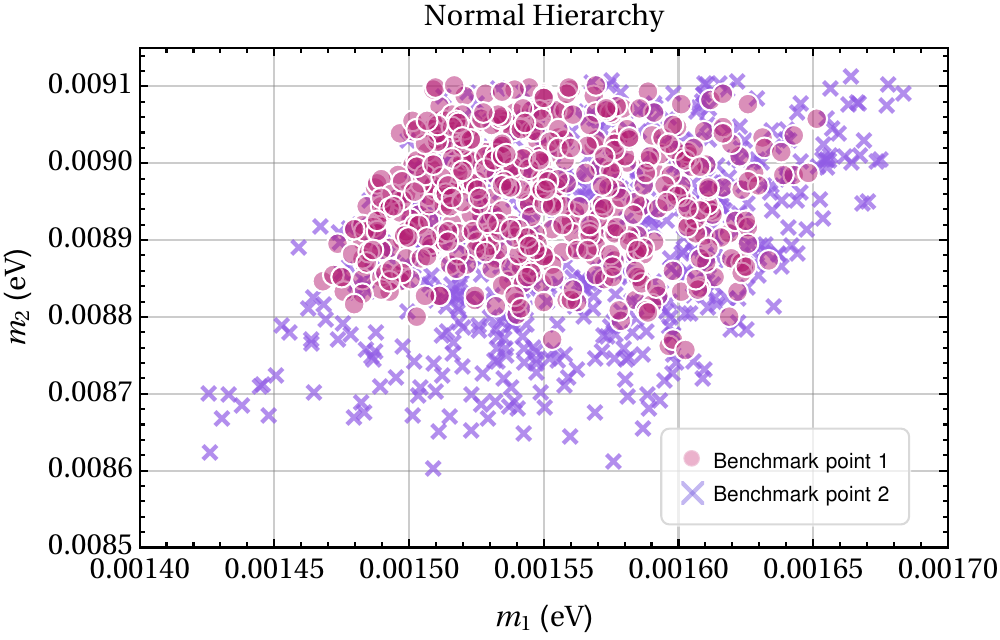}
  \caption{Plots showing the correlation among  the planes of $m_1$ and $m_2$ that are measured from our type-I seesaw model in $A_4$ modular symmetry framework. The \enquote{$\textcolor{RedViolet}{\bullet}$} and \enquote{ $\textcolor{Orchid}{\bm{\times}}$} color coded points  correspond to the  benchmark points BP1  and BP2  respectively as presented in Table. \ref{tab:BP1} and Table. \ref{tab:BP2}.}
  \label{fig5}
\end{figure}

\begin{figure}[htb!]
  \centering
    \includegraphics[width=\linewidth]{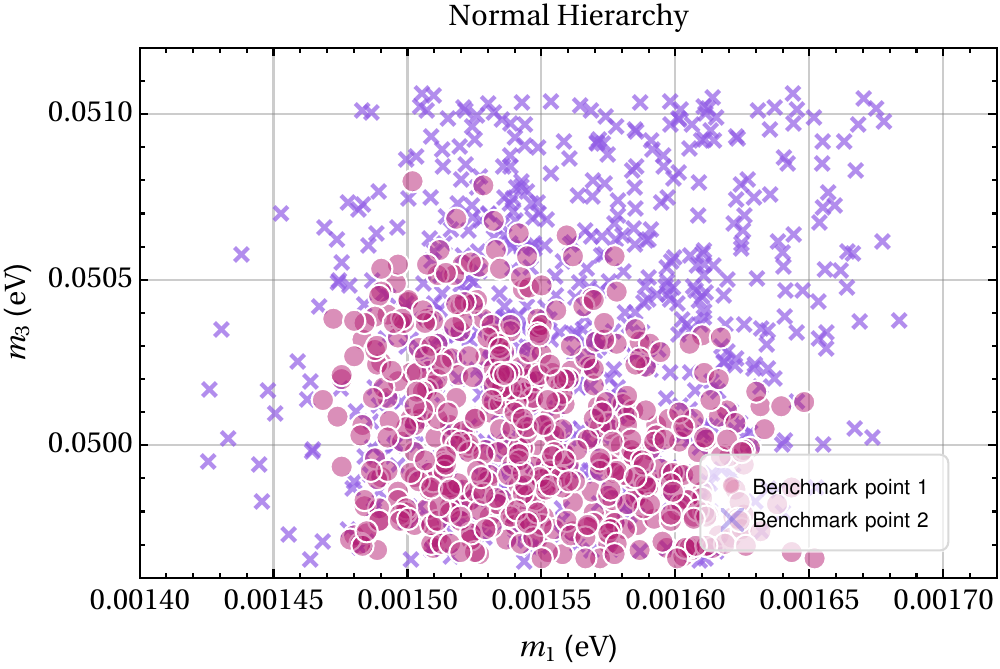}
  \caption{The plot  shows the correlation existing among $m_1$ and $m_3$ that are measured from our type-I seesaw model in $A_4$ modular symmetry framework. The \enquote{$\textcolor{RedViolet}{\bullet}$} and \enquote{ $\textcolor{Orchid}{\bm{\times}}$} color coded points  correspond to the  benchmark points BP1 and BP2 as presented in Table. \ref{tab:BP1} and Table. \ref{tab:BP2}, respectively.}
  \label{fig6}
\end{figure}

\begin{figure}[htb!]
  \centering
    \includegraphics[width=\linewidth]{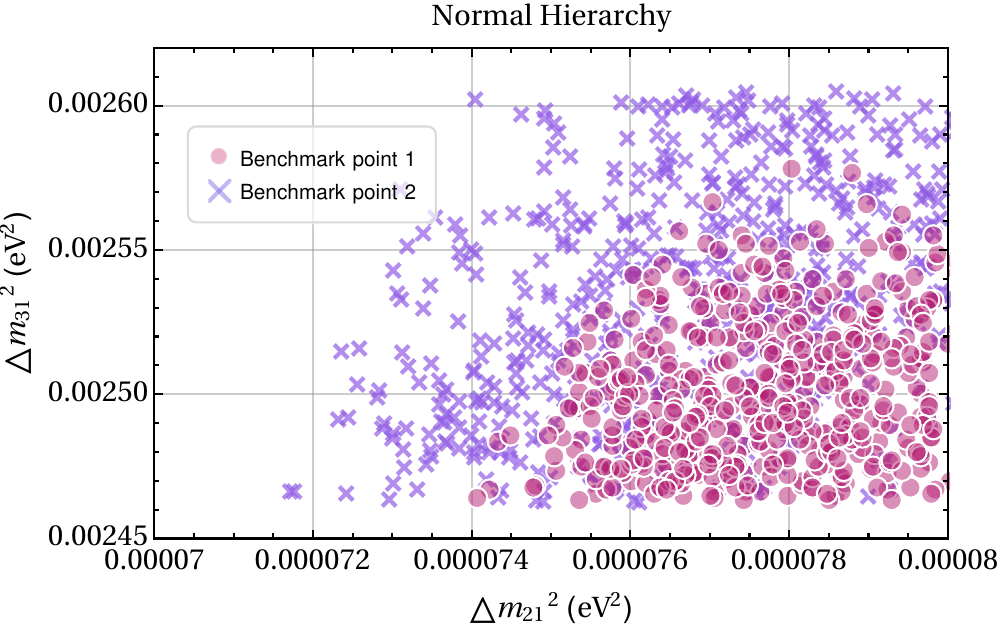}
  \caption{The $\Delta m^2_{21}$ and $\Delta m^2_{31}$ measured from our type-I seesaw invoked model shows a good correlation among them signifying the feasibility of our model and its parameter measurements. The \enquote{$\textcolor{RedViolet}{\bullet}$} and \enquote{ $\textcolor{Orchid}{\bm{\times}}$} color coded points  correspond to the  benchmark points BP1 and BP2 as presented in Table. \ref{tab:BP1} and Table. \ref{tab:BP2}, respectively.}
  \label{fig7}
\end{figure}

\begin{figure}[htb!]
  \centering
    \includegraphics[width=\linewidth]{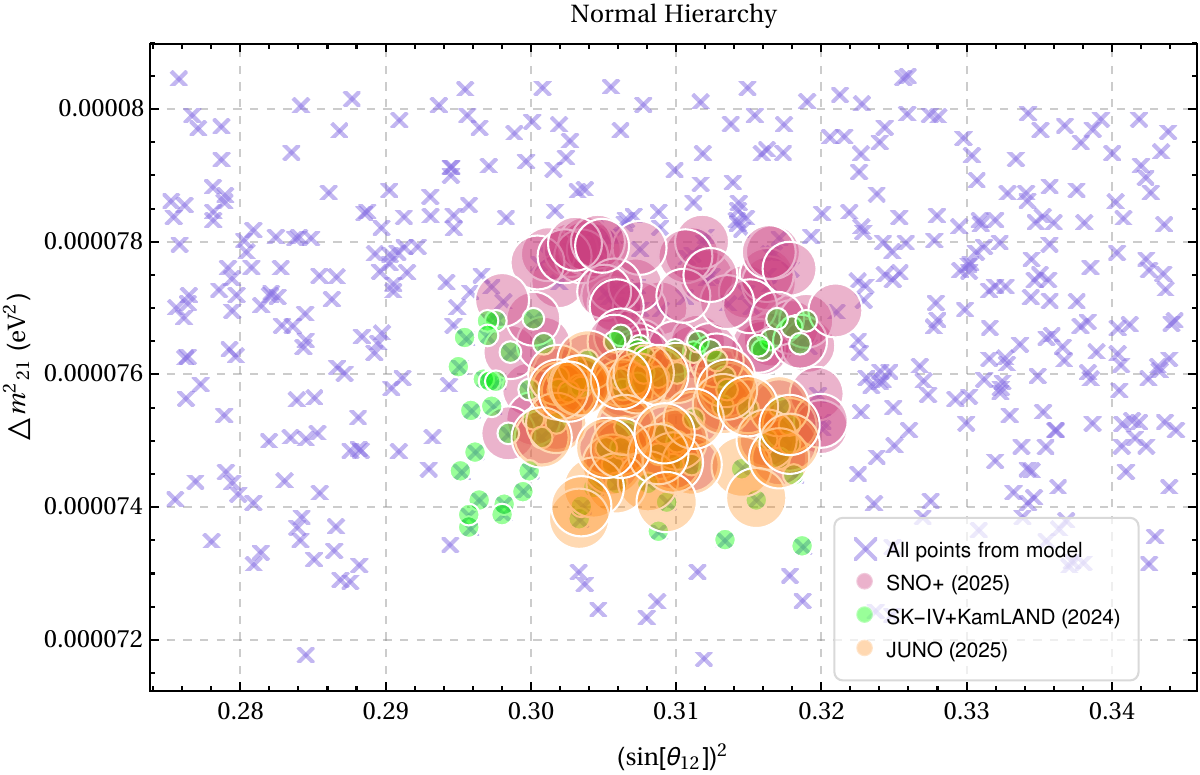}
  \caption{The plot illustates the correlation  between $ \sin^{2} \theta_{12} $  and $\Delta m^2_{21}$. The parameter $\Delta m^2_{21}$ is derived from our model and $\theta_{12} $ is sourced as input parameter from the NUFIT6.0 implying that our model parameter holds a good correlation with the experimentally measured parameters. The \enquote{$\textcolor{VioletRed}{\bullet}$}, \enquote{$\textcolor{Green}{\bullet}$}, \enquote{$\textcolor{Orange}{\bullet}$} color coded points  correspond to the correlation points that falls within the limits of SNO+ (2025), SK-IV + KamLAND (2024), JUNO(2025) and lastly the  \enquote{ $\textcolor{Violet}{\bm{\times}}$} color coded points represents all the correlation points  obtained from our model .}
  \label{fig8}
\end{figure}

In this section we have presented the correlation plots between various neutrino oscillation observables and model dependent parameters in Figs. [\ref{fig1} - \ref{fig8}]. We have presented the correlation plots between the real part of $Y^{(-2)}_{3,2}$ and the real part of the modulus parameter $\tau$ in Fig.  \ref{fig1}  as well as the correlations among real and imaginary parts of $\tau$ is shown in Fig. \ref{fig2}, within the parameter space allowed $3\sigma$. These correlations as shown in Fig. \ref{fig1} and Fig. \ref{fig2} for BP1 and BP2 in  "$\textcolor{BrickRed}{\bullet} $" and "$\textcolor{Violet}{\times}$" symbols illustrate that the modular parameter $\tau$ mainly controls the component of modular Yukawa triplets $Y^{(-2)}_{3,i}(\tau)$. Any small change in ${\rm Re}[\tau]$ or ${\rm Im}[\tau]$ results in smooth changes in triplet components. The allowed model parameters form narrow bands, demonstrating that the allowed value of $\tau$ fixed by the neutrino observables determines the modular Yukawa triplets. In Fig. \ref{fig1}, the parameter  space for the correlation plot between $Re(Y^{-2}_{3,2}) \; vs \; Re(\tau)$ has a broader region for the higher mass order of $M_D$ (BP 1) and narrower spreads from lower mass order of $M_D$ (BP 2). Since, the points are tightly clustered for BP 1 , thus it implies that the modulus parameter is more strongly constrained for BP 1. \\
On the other hand, Fig. \ref{fig2} depicts the correlation between the real and imaginary parts of the modulus parameter $\tau$. Higher order of $M_D$ masses indicate the enhanced modular space between the real and imaginary planes of the modulus parameter that transforms under the modularity condition in the upper half plane $\mathcal{H}$. This gives us the idea about the sensitivity of $M_D$ towards the modular space as derived from our model. Fig.  \ref{fig3} illustrates the robust correlation between the square of the sine of solar and reactor mixing angles within the allowed $3\sigma$ region. This correlation is crucial for the viability of our model as it impacts the parameter space of the other neutrino observables and model derived parameters.  In Fig. \ref{fig4}, we have shown the strong correlation between the square of  the sine of the solar and atmospheric mixing angles as it influences the neutrino oscillation parameters. \\

In Fig. \ref{fig5}, we illustrate the correlation of the first two light neutrino mass eigenvalues, $m_1$ and $m_2$ such that it satisfies $\Delta m^2_{12}/10^{-5} \; \in \; [6.92,\; 8.05]$ as given in NuFIT 6.0. In the first case BP1 ($M_D \sim \mathcal{O}(1\; GeV)$), the parameter space is narrower and the points are more clustered, while for BP 2 ($M_D \sim \mathcal{O}(10\; GeV)$), the parameter space has a wider space, indicating how the hierarchical structures are induced by the mass order of Dirac neutrino masses. Thus we can get a more predictive correlation space for higher $M_D$. Similarly, in Fig. \ref{fig6}, the correlation between the lightest and the heaviest masses is shown for the normal hierarchy. It is to be noted that our model predicts the hierarchy of the light neutrino masses as $m_1<m_2 <m_3$. Thus, the discrete and scattered points in Figs.  \ref{fig5} and \ref{fig6} reflect how $m_1$, $m_2$ and $m_3$ are highly constrained after the implementation of the type-I seesaw model with modular $A_4$ symmetry for both BP1 and BP2.  \\

In Fig. \ref{fig7}, we find that the solar and atmospheric  mass squared differences as obtained from our model show a very strong correlation indicating that the parameter values of $\Delta m^2_{12}$ and $\Delta m^2_{23}$ derived from our type-I seesaw model agree well its the experimental values. This hints us that our model is highly predictive and testable for ongoing and upcoming neutrino experiments. In Fig. \ref{fig8}, we have shown the correlation between the solar mixing angle $\sin^2\theta_{12}$ and the solar mass squared difference $\Delta m^2_{12}$. \\

The Jiangmen Underground Neutrino Observatory (\textsf{JUNO}~\cite{JUNO:2025gmd}) collaboration on November, 2025 reported its first data of the solar mixing angle $\theta_{12}$ and the mass-squared difference $\Delta m^2_{21}$ with unprecedented accuracy of $\Delta m^2_{21} = m_2^2 - m_1^2 = (7.50 \pm 0.12)\times 10^{-5}~\text{eV}^2$ and $\sin^2\theta_{12} = 0.3092 \pm 0.0087$ for the normal hierarchy. Apart from this, another measurement of the solar neutrino parameters was done by the SNO+ collaboration which reported  $\Delta m^2_{21} = (7.63 \pm 0.17)\times 10^{-5}\;\text{eV}^2$ and 
 $\sin^2\theta_{12}=0.310\pm0.012$ . This reactor-antineutrino data was segregated from  May~2022 to July~2025~ \cite{SNO:2025chx}. A combined best-fit neutrino oscillation parameters from recent solar experiments like Super-Kamiokande IV (2024) and KamLAND  reported $\Delta m^2_{21}/10^{-5} \approx 7.50 ^{+0.19}_{-0.18}, \; \text{and} \; \sin^2\theta_{12}= 0.307 \pm 0.012 $ \cite{Super-Kamiokande:2023jbt}. In Fig. \,\ref{fig8}, we plotted the correlation between $\sin^2\theta_{12}$ taken from NuFit 6.0 and the model derived solar mass squared differences $\Delta m^2_{12}$. We highlighted the correlation points as \enquote{$\textcolor{purple}{\times}$} for all the allowed points in our model, and as \enquote{$\textcolor{green}{\bullet}$},  \enquote{$\textcolor{red}{\bullet}$}, and \enquote{$\textcolor{orange}{\bullet}$} for points from our model that satisfy limits of SNO+ (2025), SK-IV+KamLAND (2024) and JUNO (2025) respectively.\\

For both the above mentioned benchmark points BP 1 and BP 2, the predicted value of the lightest neutrino mass is of the order of $\mathcal{O}(\text{meV})$ and the element of the effective neutrino mass $|m_{ee}|$ is predicted to be small (around a few meV), which is well below the current experimental bounds but potentially testable by next-generation and beyond ton-scale experiments. 
In both cases, our predicted value of $\delta_{CP}$ falls within the range of parameter parameters of $3\sigma$. Also, it should be noted that no data points were generated for the inverted hierarchy, which satisfies the constraints of mass-squared differences.

\begin{table}[h!]
\begin{subtable}{1\linewidth}
\centering
\begin{tabular}{|c|c|c|c|c|}
\hline
\multicolumn{5}{|c|}{Current Best Limits of $0\nu\beta\beta$ decay} \\
\hline\hline
Experiment & Isotope  &$T^{0\nu}_{1/2} \; [10^{25} \; \text{yrs}]$ & $m_{ee} \; (\text{meV})$  & Ref  \\
\hline\hline
GERDA+MJD+L200 & $^{76}\text{Ge}$ & $19$  & $ 75-200$ & \cite{LEGEND:2025jwu} \\
\hline
KamLAND-Zen 800 & $^{136}\text{Xe}$ & $38$  & $ 28-122$ & \cite{KamLAND-Zen:2024eml} \\
\hline
CUORE & $^{130}\text{Te}$ & $3.8$  & $ 70-240$ & \cite{CUORE:2024ikf} \\
\hline
\end{tabular}
\caption{}
\vspace{1 cm}
\begin{tabular}{|c|c|c|c|c|}
\hline
\multicolumn{5}{|c|}{Proposed Ton-Scale Experiments of $0\nu\beta\beta$ decay} \\
\hline\hline
Experiment & Isotope  & \multicolumn{2}{|c|}{Discovery Sensitivities of}    & Ref  \\
\cline{3-4}
 &  & $T^{0\nu}_{1/2} $ & $m_{ee} \; (\text{meV})$  &  \\
\hline\hline
CUPID & $\text{Li}_2 ^{100}\text{MoO}_4 \; \text{(crystal)}$ & $1.1 \times 10^{27}\; \text{yrs}$  & $ 12-20$ & \cite{CUPID:2022jlk}\\
\hline
LEGEND-1000 & $^{76}\text{Ge}$ & $1.3 \times 10^{28}\; \text{yrs}$  & $ 9-21$ & \cite{LEGEND:2021bnm}\\
\hline
nEXO & $^{136}\text{Xe}$ & $7.4 \times 10^{27}\; \text{yrs}$  & $ 6-27$ & \cite{nEXO:2021ujk}\\
\hline
\end{tabular}
\caption{}
\vspace{1 cm}
\begin{tabular}{|c|c|c|c|c|}
\hline
\multicolumn{5}{|c|}{Future Beyond Ton-Scale Experiments of $0\nu\beta\beta$ decay} \\
\hline\hline
Experiment & Isotope  & \multicolumn{2}{|c|}{Discovery Sensitivities of}    & Ref  \\
\cline{3-4}
 &  & $T^{0\nu}_{1/2} $ & $m_{ee} \; ({\rm meV})$  &  \\
\hline\hline
LEGEND-6000 & $^{76}\text{Ge}$ & $10^{29} \; \text{ yrs}$ & $3-6$  & \cite{LEGEND:6000}\\
\hline
ORIGIN-X & $^{136}\text{Xe}$  &  $10^{30} \; \text{ yrs}$  & $<2$ & \cite{Heffner}\\
\hline
THEIA & $^{130}\text{Te}(^{136}\text{Xe})$ & $> 1.1 (2.0) \times 10^{28}\; \text{yrs}$  & $< 6.3 (5.6)$  & \cite{Kaptanoglu}\\
 &  & $90\% \; \text{C.L.} \; \text{for }\;\text{Te (Xe)}$  &  &  \\
\hline
\end{tabular}
\caption{}
\end{subtable}
\caption{Current limits of $0\nu\beta \beta$ decay, projected sensitivities of the proposed ton-scale experiments and the future beyond ton-scale experiments presented in Table 7(a), 7(b), 7(c), respectively. If somehow no discovery is made at the ton-scale experiments or its predecessors then their would be next-next generation experiments whose target would be to achieve $m_{ee} \sim 10 \; \mbox{meV}$ or smaller. The main goal of  the experiments like LEGEND-6000 and ORIGIN-X would be to expedite and cover the parameter space of $m_1-m_{ee}$ for normal ordering beyond $m_{ee} \sim 10 \; \mbox{meV}$.}
\label{tab:nubb}
\end{table}

\subsection{Implications to Neutrinoless Double Beta Decay:}
One of the striking features of Majorana neutrinos is that it can be probed via lepton number violation in neutrinoless double decay. We have briefly discussed here the allowed range of the input parameters in the light neutrino mass matrix and their implications for neutrinoless double beta decay. Such a rare process is possible when two neutrons simultaneously decay into 2 protons and 2 electrons but without emission of any neutrinos. 

The $0\nu\beta\beta$ is represented by:
\begin{equation}
	0\nu \beta \beta
	\, : \, 
	(A, \, Z) \rightarrow (A, \, Z + 2)^{++} + 2 e^- 
	\, . 
\end{equation}
In the present scenario, $0\nu\beta\beta$ decay can be induced by exchange of light Majorana neutrinos or heavy Majorana $RHN$. However, the high-scale right-handed neutrinos considered for thermal leptogenesis give negligible contributions to the decay rate of $0\nu\beta\beta$. 

\begin{figure}[t]
    \centering
    \hspace{-0.5cm}
    \includegraphics[width=1.1\linewidth]{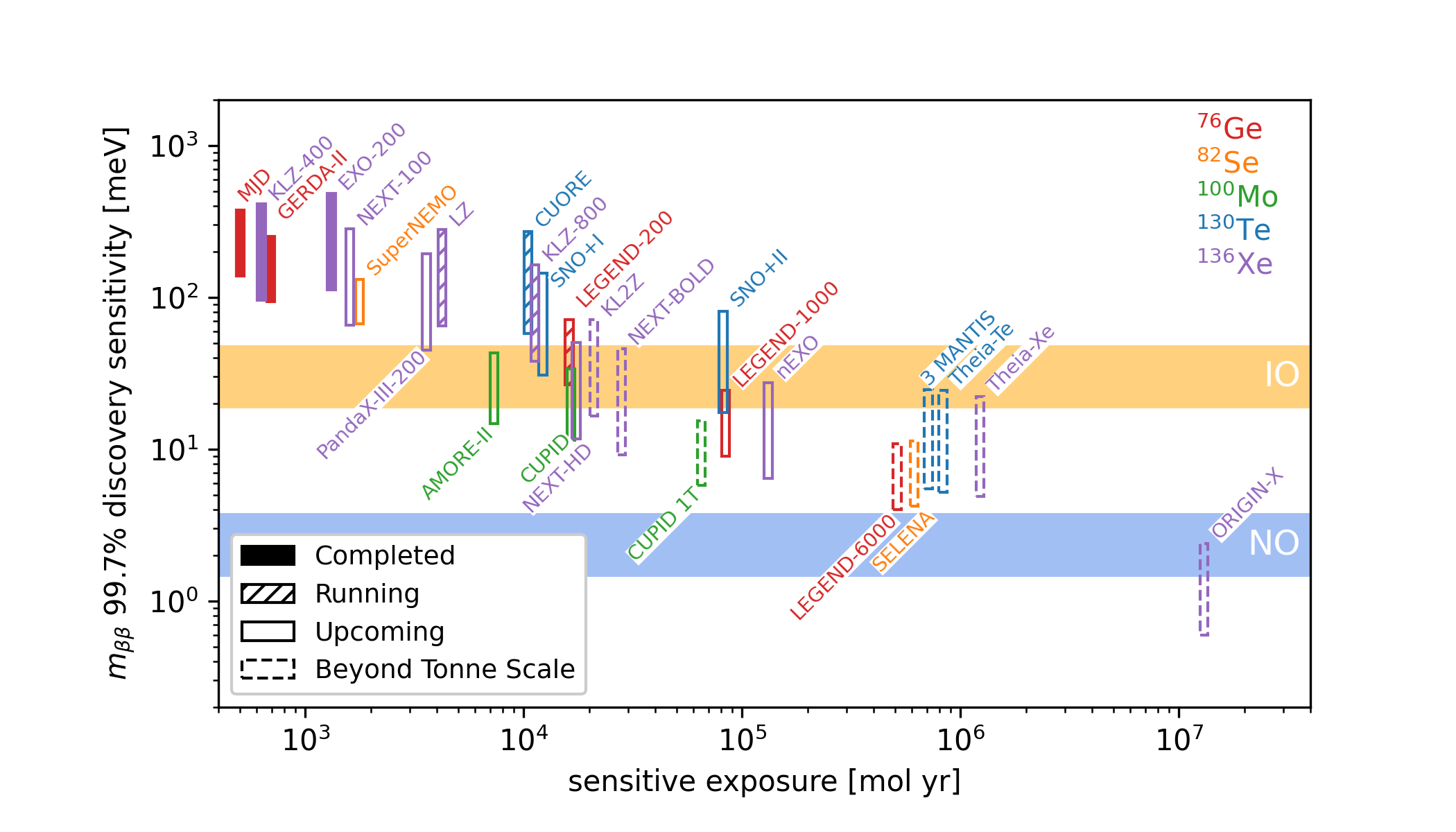}
    \caption{Projected experimental sensitivity on effective Majorana mass parameter $m_{ee}$ for various neutrinoless double beta decay experiments. This image is initially produced and discussed in ref~\cite{Anker:2024xfz} and utilized here to demonstrate the sensitivity of various neutrinoless double beta decay on the effective Majorana mass parameter $m_{ee}$. It can be seen that the Origin-X is the most promising beyond ton-scale experiment to probe the parameter space of $m_{ee}$ beyond 10 meV for normal ordering.}
    \label{fig:expts}
\end{figure}
\begin{figure}[t!]
\centering
	\includegraphics[width=0.65\textwidth]{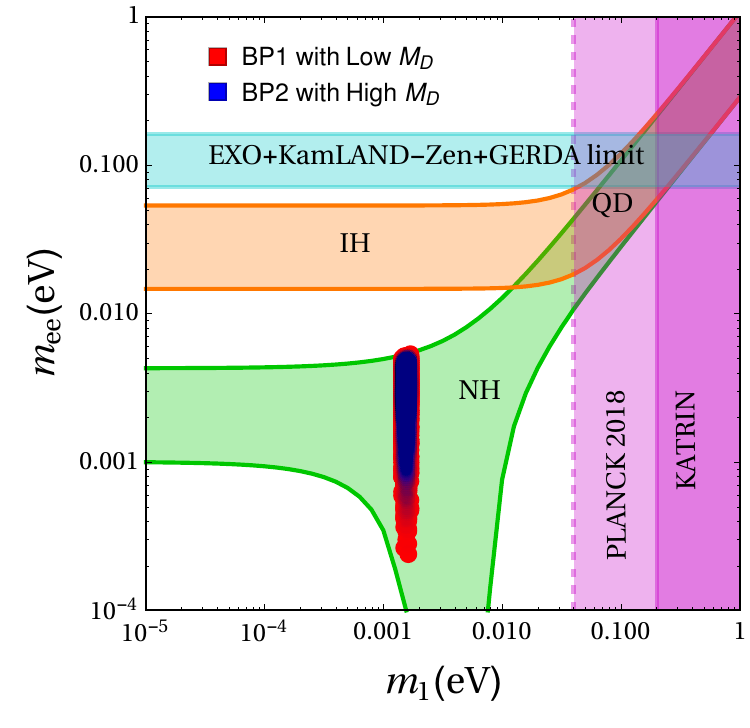}
\caption{
Plot for effective Majorana mass $m_{ee} (\mbox{eV})$ with variation of lightest neutrino mass $m_1$ for normal hierarchy. The green and red color band shows the expected behavior of $m_{ee}$ by varying neutrino oscillation parameters within $3-\sigma$ range. The vertical lines and the corresponding shaded areas represent the limit on lightest neutrino masses coming from cosmological data (PLANCK 2018) and the KATRIN experiment. The horizontal shaded region presents the translated experimental limit on the lightest neutrino mass from the half-life taken from GERDA and EXO+KamLAND-Zen experiments. The model predictions of $m_{ee}$ are estimated for both benchmark points, displayed in red and blue color points.}
\label{fig:nubb}
\end{figure}

The limit on the effective parameter of the Majorana mass is derived by translating the existing lower limit on the half-life of $0\nu\beta\beta$. The inverse half-life of this $0\nu\beta\beta$ decay process can be expressed in terms of the effective Majorana mass parameter, the nuclear matrix element, and the phase-space factor as follows:
\begin{equation}
    \left[
\left( T_{1/2}^{0\nu} \right)^{-1}_{(A,Z)} 
= G^{0\nu}_{(A,Z)} \, \left| \frac{M^{0\nu}}{m_e} \right|^2 \, \left| m_{\beta\beta} \right|^2
\right]
\end{equation}
where effective Majorana mass which is a measure of $LNV$ can be written down in terms of lightest neutrino mass and oscillation parameters as: 
\begin{equation}
    |\langle m_{\beta\beta}\rangle | = \vert \sum\limits_{i=1} m_i U^2_{ei} \vert = \vert c^2_{12} c^2_{13}e^{2i\alpha}m_1 + c^{2}_{13}s^2_{12}e^{2i\beta}m_2 +s^2_{13}e^{-2i\delta}m_3 \vert
    \end{equation}

The variation of effective Majorana mass parameter $m_{ee}$ with lightest neutrino mass $m_1$ ($m_3$) for normal ordering (inverted ordering) is displayed in Fig. \ref{fig:nubb}. The horizontal bands are for the well-established current experimental limit on neutrinoless double beta decay as given in Table [\ref{tab:0nbb_limits}]. The vertical magenta color bands indicate bounds from KATRIN and PLANCK data on the absolute neutrino mass scale. 
\\
The shaded regions in green and beige colors as shown in the figure, corresponds to $NH$ and $IH$ pattern. This depicts how $m_{\beta\beta}$ varies with light neutrino masses while changing the other oscillation parameters in their $3\sigma$ range. However, this tightly constrains the $A_4$ modular seesaw-driven parameters that are considered in our model. As a result, the model parameters yield sub meV range for $m_{\beta\beta}$ as shown in red color. The red-colored data points indicate the points for BP1 with Low $M_D$ and the blue-colored data points indicate the points for BP2 with High $M_D$. The current bounds on the lightest neutrino mass are ($m_\nu < 0.45$ eV at $90\%$  CL) derived from the KATRIN experiment ~\cite{KATRIN:2024cdt}.

On the other hand, the $0\nu\beta\beta$ contributions arising from light neutrinos assuming a quasi-degenerate pattern can saturate the present experimental limit, but they are unfavorable by cosmology data. However, the assumed NH mass hierarchy results in an effective Majorana mass parameter in the range of few meV, which is beyond the scope of the currently planned experiment but may be probed in the near-future multi-term scale detector. Alternatively, any near-future confirmation of this rare event would require new physics.

\begin{table}[h!]
\begin{center}
	\begin{tabular}{c|c|c|c}
	Isotope & $T_{1/2}^{0 \nu}~[10^{25}\text{ yrs}]$ & $m_\text{eff}^{0 \nu}~[\text{eV}]$ & Collaboration \\
	\hline
	$^{76}$Ge	& $> 2.1$	& $< (0.2 - 0.4)$ 	& GERDA~\cite{Agostini:2013mzu} 			\\
	$^{136}$Xe	& $> 1.6$	& $< (0.14 - 0.38)$ 	& EXO~\cite{Auger:2012ar} 				\\
	$^{136}$Xe	& $> 1.9$	& n/a 				& KamLAND-Zen~\cite{Gando:2012zm} 		\\
	$^{136}$Xe	& $> 3.6$	& $< (0.12 - 0.25)$ 	& EXO + KamLAND-Zen combined~\cite{Gando:2012zm} 	
	\end{tabular}
\caption{The current lower limits on half-life $T_{1/2}^{0 \nu}$ and the translated upper limits of effective Majorana mass parameter $m_\text{eff}^{0 \nu}$ in the $0\nu\beta \beta$ decay experiments like GERDA, EXO, KamLAND-Zen are presented here. The range for the $m_\text{eff}^{0 \nu} $ arises from the NME calculated by different methods.}
\label{tab:0nbb_limits}
\end{center}
\end{table}

\section{Leptogenesis within type-I seesaw framework}
\label{sec:lepto}
Leptogenesis has been an elegant framework for generating the observed baryon asymmetry of the Universe, which cannot be addressed within SM. The key idea of leptogenesis is to first create an asymmetry in the lepton sector, and subsequently the created lepton asymmetry becomes a net baryon asymmetry through anomalous $B+L$ violating sphaleron processes~\cite{Klinkhamer:1984di, Arnold:1987mh, Kuzmin:1985mm, Rubakov:1996vz}. In the original proposal of leptogenesis~\cite{Fukugita:1986hr}, the CP-violating and lepton number violating out of equilibrium decays of heavy Majorana right-handed neutrinos introduced within the Type-I seesaw mechanism. Many variants of leptogenesis~\cite{Langacker:1986rj, Luty:1992un, Murayama:1992ua, Acker:1993vx, Buchmuller:1996pa} have been explored, and thermal leptogenesis is one of its kind, where the typical mass of right-handed neutrinos is considered at a high scale of around $10^{10}$~ GeV  or more. Depending on how we account for flavor parts in the determination of CP-asymmetry, there are two classes of thermal leptogenesis: unflavored and flavored type, and we limit ourselves to unflavored leptogenesis in the rest of the analysis. 

 Estimation of CP-asymmetry is crucially dependent on the exact structure of the Dirac neutrino mass matrix $M_D$ and the mass eigenvalues $M_i$ of right-handed neutrinos.  However, within the type-I seesaw, the Dirac mass matrix ($M_D$) or Yukawa coupling ($Y_D$) and the Majorana mass matrix ($M_R$) for right-handed neutrinos contain many free parameters; therefore, the high-scale CP asymmetries relevant for the leptogenesis are mostly unconstrained by low-energy data. Alternatively, a flavor symmetry, such as modular symmetry $A_4$, can significantly reduce the free model parameters by useful relations among the Yukawa matrix $Y_D$ and $M_R$ elements. Thus, leptogenesis predictions with a small subset of model parameters can be easily verified with low-energy observables, and we can get nice correlations between the low-energy CP-violating phase and high-scale phases involved in leptogenesis. 

Let us briefly discuss the type-I seesaw mechanism and its connection to leptogenesis. The seesaw mechanism provides a natural framework for understanding the tiny neutrino masses by adding right-handed neutrinos to the SM~\cite{Minkowski:1977sc, Gell-Mann:1979vob,  Yanagida:1979as, Mohapatra:1979ia, Schechter:1980gr, Mohapatra:1980yp, Schechter:1981cv} and, at the same time, a dynamical origin for the baryon asymmetry of the Universe through leptogenesis. The relevant Lagrangian contains the heavy right-handed Majorana neutrinos $N_{R_i}$, the Standard Model lepton doublets $L_\alpha$, and the Higgs doublet $H$:
\begin{equation}
\mathcal{L} \supset - \overline{L}_\alpha \, [Y_D]_{\alpha i} \, \tilde{H} \, N_{Ri} - \frac{1}{2} \overline{N_{Ri}^c} (M_R)_{ij} N_{Rj} + \text{h.c.},
\end{equation}
where $Y_D$ is the Dirac Yukawa matrix and $M_R$ is the Majorana mass matrix of the right-handed neutrinos. These interaction terms provide all necessary  Sakharov conditions for realizing leptogenesis:
\begin{enumerate}
\item[1.] \textbf{Lepton number violation:} The presence of Majorana mass terms for the right-handed neutrinos explicitly violates lepton number.
\item[2.] \textbf{CP violation:} Complex phases in the neutrino Yukawa couplings provide the required source of CP asymmetry in heavy neutrino decays.
\item[3.] \textbf{Departure from thermal equilibrium:} Out-of-equilibrium decays occur when the decay rate ($\Gamma_D$) of the lightest RHN ($N_1$) is smaller than the Hubble expansion rate of the Universe ($H(T)$) at temperature $T=M_1$. This requirement may be expressed schematically as
\begin{equation}
\Gamma_D \leq \textbf{H}(T= M_1) , .
\label{eq:dr}
\end{equation}
The decay rate ($\Gamma_D$) and Hubble expansion rate ($H(T)$) are expressed in terms of lightest RHN mass as follows 
\begin{equation}
\Gamma_{D} = \frac{(Y^\dagger Y){ii},M_i}{8\pi}, \hspace{1cm}
\textbf{H}(T) = \frac{2}{3},\sqrt{\frac{g* \pi^3}{5}} ,\frac{T^2}{M_{\rm Pl}}
\end{equation}
with $M_{\rm Pl}=1.22\times 10^{19}$ GeV is the Planck mass and $g_*$ (=106.75 for SM ) represents total number of relativistic degrees of freedom excluding RHNs contributing to the energy density of the universe. Within \emph{weak washout regime} when Eq.~(\ref{eq:dr}) is satisfied, the heavy RHNs can not thermalize completely with the cosmic plasma. 
In this case, even a slight CP-violating asymmetry in the decays will survive and contribute to the lepton asymmetry, making leptogenesis practically unavoidable.

On the other hand, if 
$\Gamma_D > H(T= M_1)$, the system lies in what is termed the \emph{strong washout regime}. Contrary to naive expectation, this does not automatically eliminate the possibility of generating the observed baryon asymmetry. Instead, the outcome depends on a more delicate balance between decays, inverse decays, and scattering processes. The detailed studies requires solving the coupled Boltzmann equations that track the lightest RHN $N_1$ number densities and lepton asymmetries. 
It has been demonstrated that the strong washout scenario often yields robust predictions for the final baryon asymmetry, since the system tends to lose sensitivity to the precise initial conditions~\cite{Davidson:2008bu, Joshipura:2001ya}. 
\end{enumerate}
It was pointed out in initial works~\cite{Flanz:1994yx, Flanz:1996fb, Pilaftsis:1997jf, Pilaftsis:1997dr, Roulet:1997xa, Buchmuller:1997yu, Flanz:1998kr, Hambye:2004jf, Pilaftsis:2003gt} that there are two sources of CP-violation: 1) interference of tree level diagram with one loop vertex diagram and 2) interference between tree level with one loop self energy diagram. In the assumed mass hierarchy $M_1 \ll M_2\simeq M_3$, the self-energy diagram gives a negligible contribution while the vertex diagram dominates. The CP asymmetry in the decays of RHNs $N_i$ is defined as:
\begin{equation}
		\epsilon_{i\alpha} \equiv \frac{\Gamma\left(N_i \rightarrow L H  \right)-\Gamma\left(N_i \rightarrow \overline{L} H^*\right)}{\Gamma\left(N_i\rightarrow L H \right)+\Gamma\left(N_i \rightarrow \overline{L} H^*\right)}\equiv \frac{\Delta \Gamma_{N_i}^\alpha}{\Gamma_{D}}
	\end{equation}
The total decay rate $\Gamma_{D}$ for the decay of right handed neutrinos to leptons and Higgs is defined as
\begin{equation}\label{GammaTotal}
		\begin{split}
			\Gamma_{D} & =\Gamma(N_{i}\rightarrow
			L H)+\Gamma(N_{i}\rightarrow
			\overline{L} H^* )\\
		\end{split}
\end{equation}
One can write the Dirac mass matrix as an effective Yukawa coupling multiplied by the electroweak VEV, $M_D=v\, Y_D$. The CP-asymmetry arising from the interference between tree and one-loop contributions \cite{Covi:1996wh}  of the CP-violating and lepton-number-violating decays of the lightest RHN in terms of  $Y_D$ or $M_D$ and masses of RHNs as
\begin{equation}
	\begin{aligned}
		\epsilon_{i \alpha}= & \frac{1}{8 \pi} \frac{1}{\left(Y^{\dagger}_D Y_D\right)_{i i}} \sum_{j \neq i} \operatorname{Im}\left[\left(Y^{\dagger}_D Y_D\right)_{j i} [Y_D]_{\alpha i} [Y_D]_{\alpha j}^*\right] g\left(\frac{M_j^2}{M_i^2}\right) \\
		& +\frac{1}{8 \pi} \frac{1}{\left(Y^{\dagger}_D Y_D\right)_{i i}} \sum_{j \neq i} \operatorname{Im}\left[\left(Y^{\dagger}_D Y_D\right)_{i j} [Y_D]_{\alpha i} [Y_D]_{\alpha j}^*\right] \frac{M_i^2}{M_i^2-M_j^2},
	\end{aligned}
\end{equation}
where $g(x_j)$ with $x_j=M^2_j/M_1$ is the loop function
\begin{equation}
	g(x_j)=\sqrt{x_j}\left[\frac{1}{1-x_j}+1-(1+x_j) \ln \left(\frac{1+x_j}{x}\right)\right]
\end{equation} 
Within type-I seesaw mass hierarchy $M_R \gg M_D$ and with hierarchy among RHN masses $M_1 \ll M_2\simeq M_3$, the simplified expression for CP-asymmetry within unflavor approximation is as follows
		\begin{eqnarray}
			\epsilon_1&\approx& -\frac{3 M_{1}}{16\pi(Y_D^\dagger Y_D)_{11}}\left[\frac{Im[(Y_D^\dagger Y_D)^2_{21}]}{M_{2}}+\frac{Im[(Y_D^\dagger Y_D)^2_{31}]}{M_{N_3}}\right]
		 \nonumber \\
	&=& -\frac{3 M_{1}}{16\pi v^2(M_D^\dagger M_D)_{11}}\left[\frac{Im[(M_D^\dagger M_D)^2_{21}]}{M_{2}}+\frac{Im[(M_D^\dagger M_D)^2_{31}]}{M_{3}}\right].
			\label{epsilon1MD}
	\end{eqnarray}  
Thus, the required value of CP-asymmetry depends upon the Dirac neutrino mass matrix $M_D$ structure and mass eigenvalues of right-handed Majorana neutrinos. The important feature of the present $A_4$
modular symmetry within Type-I seesaw results in a simplified structure of $M_D$ and $M_R$ leading to predictive light neutrino masses and mixings. In the present work, we wish to investigate a specific non-trivial RHN mass texture and its impact on the CP-violating parameter relevant for leptogenesis consistent with neutrino masses and mixings.


\subsection{CP-asymmetry in Type-I Seesaw and Role of \texorpdfstring{$A_4$}{A4} Modular Symmetry}

The magnitude of the CP-violating asymmetry is controlled by the texture of the Dirac neutrino mass matrix $M_D$ and the eigenvalue spectrum of the heavy Majorana neutrinos $M_1, M_2, M_3$. A notable outcome of implementing the present $A_4$ modular symmetry in the Type-I seesaw framework naturally yields a simplified form of $M_D$ and $M_R$. As a result, the light neutrino mass spectrum and their mixing pattern are explained with minimal model parameters as an outcome of this modular $A_4$ symmetry. The derived structure of the Dirac neutrino mass matrix is given by
\begin{eqnarray}
M_D=\left(
\begin{array}{ccc}
  -11.14-0.405\,i & -7.723+3.973\,i   & -1.729+1.709\,i \\
 -6.768+3.482\,i & -2.896+3.482\,i & 7.591-0.276 i \\
 -2.538+2.509\,i & 12.71-0.463\,i    & -4.610+2.372\,i
\end{array}
\right) \quad \mbox{GeV}
\label{MD1} 
\end{eqnarray}
and the right-handed neutrino mass matrix takes the non-diagonal form
\begin{equation}
M_R =
\begin{pmatrix}
a & 0 & 0 \\
0 & 0 & b \\
0 & b & 0
\end{pmatrix}, \qquad b = 10a, \quad a \in \mathbb{R}.
\end{equation}
This structure leads to a $2\times2$ sub-block mixing in the $(2,3)$ sector, giving eigenvalues
\begin{equation}
M_1 = a, \qquad M_2 = -b, \qquad M_3 = b.
\end{equation}
The negative eigenvalue can be positive via a field redefinition $N_2 \rightarrow i N_2$, implemented through a diagonal phase matrix $P = \text{diag}(1, i, 1)$. The diagonalization thus proceeds as
\begin{equation}
M_R^{\text{diag}} = \mathcal{P}^T O^T M_R O \mathcal{P} = \text{diag}(M_1, M_2, M_3),
\end{equation}
Where $O$ represents an orthogonal matrix that diagonalizes the real part of RHN mass matrix $M_R$, and $\mathcal{K}$ ensures all eigenvalues are positive. The successful leptogenesis assumes a diagonal RHN mass matrix with real and positive mass eigenvalues—alternatively, the modular $A_4$ symmetry results in a non-diagonal RHN mass matrix structure.  Thus, we must go to a different basis where RHNs can be made diagonal by this rotation matrix $ U_R=O\mathcal {P}$. This change of basis will also modify the structure of the Dirac neutrino mass matrix $M_D$ to $\widehat{M}_D = M_D O \mathcal{P}$. From now onward, we will limit our numerical estimation of CP-asymmetry and other parameters in terms of $\widehat{M}_D$. As a result, the modified structure of the Dirac Yukawa matrix becomes
\begin{equation}
\widehat{Y}_D = Y_D U_R = Y_D O \mathcal{P} \quad \quad \mbox{or,} 
\widehat{M}_D = M_D O \mathcal{P}
\end{equation}
which directly enters into the expression of the CP-asymmetry parameter through the combination $\widehat{Y}_D^\dagger \widehat{Y}_D$. The modified expression for CP-asymmetry is given by 
\begin{eqnarray}
			\epsilon_1&\approx& -\frac{3 M_{1}}{16\pi(\widehat{Y}_D^\dagger \widehat{Y}_D})\left[\frac{Im[(\widehat{Y}_D^\dagger \widehat{Y}_D)^2_{21}]}{M_{2}}+\frac{Im[(\widehat{Y}_D^\dagger \widehat{Y}_D)^2_{31}]}{M_{3}}\right]
		 \nonumber \\
	&=& -\frac{3 M_{1}}{16\pi v^2(\widehat{M}_D^\dagger \widehat{M}_D)_{11}}\left[\frac{Im[(\widehat{M}_D^\dagger \widehat{M}_D)^2_{21}]}{M_{2}}+\frac{Im[(\widehat{M}_D^\dagger \widehat{M}_D)^2_{31}]}{M_{3}}\right]
			\label{epsilon1MD2} \nonumber \\
    &\simeq & -\frac{3}{8\pi v^2 (\widehat{M}_D^\dagger \widehat{M}_D)_{11}} 
    \frac{M_1}{M_{k}} \left[Im[(\widehat{M}_D^\dagger \widehat{M}_D)^2_{21}]+Im[(\widehat{M}_D^\dagger \widehat{M}_D)^2_{31}]\right].
			\label{epsilon1MD1}
\end{eqnarray}
Here, the index k can be 2 or 3 as $M_2\simeq M_3$ is considered.  

 This departure from equilibrium is quantified by the decay parameter K as
\begin{equation}
    K = \frac{\Gamma_N}{\textbf{H}(T = M_{N_1})} = \frac{\widetilde{m_1}}{m^*},
\end{equation}
where the effective and equilibrium neutrino mass parameters are given as 
\begin{equation}
    \widetilde{m_1} = \frac{(\widetilde{Y}^\dagger_D \widetilde{Y})_{11}v^2}{M_{1}},\quad 
    m^* = \left. 8\pi\frac{ v^2}{M_{1}} \right|_{T = M_{1}}\,.
\end{equation}
We make comparison between two regimes based on the value of the $K$ as $K \ll 1$, and $K \gg 1$, corresponding to the weak, and strong washout regimes. In our analysis, using the best-fit values of the input model parameters associated with the modular symmetry, we find that the dominant contribution to the lepton asymmetry arises from the decays and inverse decays of the heavy right-handed neutrinos in the strong washout regimes.

We estimated the numerical values of the decay rate $\Gamma_D$, Hubble expansion rate $H(T=M_1)$, and CP-asymmetry for two different choices of input model parameters, including the important complex modulus parameter $\tau$. These estimated values are presented in  Table~\ref{tab:cpasy}, highlighting the difference between both weak and strong washout behaviors.   For each case, the chosen set of prefactors $(\alpha_D, \beta_D, \gamma_D)$ associated with the Yukawa couplings are fixed at $(0.2,\,0.3,\,0.4)$. We fixed right-handed neutrino masses as input model parameters with the lightest state $M_1$ falling within the range of $10^{12}$–$10^{14}\,\text{GeV}$ and $ M_2\simeq M_3$, which are approximately an order of magnitude larger. The estimated value of the washout parameter $K$ remains almost the same for both cases, at $88$, indicating strong washout in both benchmark scenarios. 

\begin{table}[h!]
\centering
\begin{tabular}{|c|c|c|c|c|c|}
\hline 
Complex modulus \boldmath$\tau$ &
($\pmb{\alpha_D}$, $\pmb{\beta_D}$, $\pmb{\gamma_D}$) & \boldmath$M_1$ (GeV) & \boldmath$M_2 \simeq M_3$ (GeV) & \boldmath$K = \Gamma_1/H$ & \pmb{$\epsilon_{cp}$}  \\
\hline
$0.20 + 1.22i$ & $0.2$, $0.3$, $0.4$ & $3.7 \times 10^{12}$ & $2.35 \times 10^{13}$  & $88.6$  & $10^{-6}$\\
\hline
$-0.17 + 1.23i$ & $0.2$, $0.3$,$0.4$ & $8.68 \times 10^{13}$ & $6.67 \times 10^{14}$  & $88.7$ & $10^{-8}$ \\
\hline
\end{tabular}
\caption{Benchmark points used to generate the lepton asymmetry in the model.}
\label{tab:cpasy}
\end{table}

We find that the CP asymmetry parameter $\epsilon_{cp}$ is sensitive to the choice of modulus, varying from $10^{-6}$ for $\tau = 0.20 + 1.22i$ to $10^{-8}$ for $\tau = -0.17 + 1.23 i$.  These representative sets of parameters enable us to investigate the impact of heavy RHNs on leptogenesis and the corresponding CP asymmetry. Within strong washout regimes, the predicted values of CP asymmetry is found to be sensitive to the modulus parameter $\tau$ and the estimated value of $\varepsilon_1$ varies from $10^{-6}$ for $\tau = 0.20+1.22i$ to $10^{-8}$ for $\tau = -0.17+ 1.23 i $ .

The generated lepton asymmetry, within modular $A_4$ symmetric type-I seesaw  mechanism, is then partially converted into a baryon asymmetry via well known electroweak sphaleron transitions as follows
\begin{equation} 
Y_{B} = c_s Y_{B-L}, \qquad c_s = \frac{8N_f + 4N_H}{22N_f + 13N_H}, \end{equation} 
where $c_s$ is the conversion factor which is measure of how much fraction of the lepton asymmetry that has been already created and converted into final baryon asymmetry. Using $N_f = 3$ and $N_H = 1$, the final baryon asymmetry can be rewritten as 
\begin{equation}
Y_{B} \simeq c_s \kappa\, \frac{\epsilon_1}{g^*} = \frac{28}{79} Y_{B-L}. 
\end{equation}
The other important dilution factor $\kappa$ accounts for washout processes and inverse decays that can erase the already produced asymmetry. It is parameterized as follows~\cite{kolb_turner_1990}: 
\begin{eqnarray}
-\kappa &\simeq &  \sqrt{0.1K} \text{exp}[-4/(3(0.1K)^{0.25})], \;\; \text{for} \; K  \ge 10^6 \nonumber \\
&\simeq & \frac{0.3}{K (\ln K)^{0.6}}, \;\; \text{for} \; 10 \le K \le 10^6 \nonumber \\
&\simeq & \frac{1}{2\sqrt{K^2+9}},  \;\; \text{for} \; 0 \le K \le 10.
\end{eqnarray}
Thus, the estimated final baryon asymmetry is consistent with the WMAP results for baryon asymmetry:  
\begin{equation} Y_{B} \equiv \frac{n_B-n_{\bar{B}}}{s}\bigg|_0 = (8.75 \pm 0.23) \times 10^{-11}. 
\end{equation}


\subsection{Boltzmann Equations}
We will examine the evolution of lepton asymmetry and quantify the final baryon asymmetry via the survived lepton asymmetry as the universe cools down gradually via the well-known Boltzmann equations (BEs) by considering the decays and inverse decays of the lightest neutrino $N_1$ only. In our earlier discussion, it has been demonstrated that one can get the required value of CP-asymmetry parameter $|\epsilon_1|$ and thereby, generate a correct value of baryon to photon ratio $Y_B\simeq 10^{-10}$. However, it is important to examine the evolution of both $N_1$ density ($Y_{N_1}$) and already created $B-L$ asymmetry ($Y_{\Delta L}$) in the lepton sector. The reason for tracking the evolution of these two quantities is that $N_1$ density might be insufficient or processes involving SM particles can overpopulate RHNs. At the same time, inverse decays $L H  \to N_1$ might be too strong and wash away any lepton asymmetry already created due to decay of $N_1$'s. Our analysis is limited to the single-flavor regime where the relevant coupled Boltzmann equations describing the evolution of $N_1$ density ($Y_{N_1}$ ) and the $B-L$ asymmetry ($Y_{\Delta L}$) are as follows:
\begin{equation}
	\begin{aligned}\label{BEs}
		\frac{dY_{N_1}}{dz} & =-D_1(z) \bigg[Y_{N_1}(z) - Y_{N_1}^{eq}(z) \bigg] \\
		\frac{dY_{\Delta L}}{dz} & = \epsilon_1 D_1(z) \bigg[Y_{N_1}(z) - Y_{N_1}^{eq}(z) \bigg]-W_{1}(z)\, Y_{\Delta L}(z)
	\end{aligned}
\end{equation}
Here, we introduce the abundances parameter ($Y_i=\frac{n_i}{s}$ with i=$N_1, \Delta L$) as ratios of the particle densities $n_i=\int d^3p f_i$ to the entropy density $s=\displaystyle{\frac{2\pi^2}{45}} g_* T^3$ in order to scale out the effect of the expansion of the Universe. Other key parameters used in BEs are 1) $z=\frac{M_1}{T}$ is a dimensionless parameter varying inversely with the temperature of the Universe, 2) $Y_{\Delta L}=2(Y_\ell-Y_{\bar{\ell}})$. The decay and washout terms are, respectively, given by 
\begin{eqnarray}
D_1(z) &=& \frac{z\,\gamma_{N_1}}{s\,H(M_1)} =
z\,K_1\,\frac{{\cal K}_1(z)}{{\cal K}_2(z)},\;\;\;\;\;\;\;\; W_1(z) =
\frac{1}{2}D_1(z) \frac{Y_{N_1}^{eq}(z)}{Y_{\ell}^{eq}},
\end{eqnarray}
with ${\cal K}_n$ the n-th order modified Bessel function of second kind.
The equilibrium abundances for lightest RHN $N_1$ and leptons $\ell$ are given by
\begin{eqnarray}
Y_{N}^{\mathrm{eq}}(z)=\frac{45}{2\pi^4 g_*}z^2 \mathcal{K}_2(z)\, \quad Y_{\ell}^{\mathrm{eq}}(z)=\frac{15}{4\pi^2 g_*}				
\end{eqnarray}
\begin{figure}[t!]
\centering
\includegraphics[width=0.9 \textwidth]{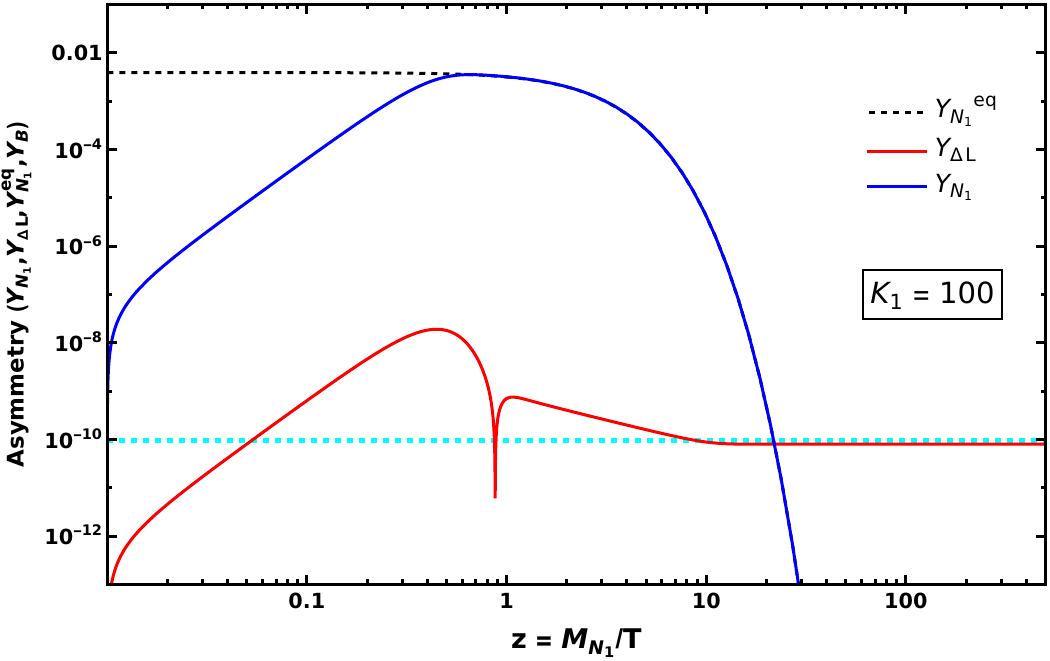}
\caption{Evolution plot for comoving number densities with the variation of $z=T/M_1$ in the strong washout regime ($K_1 = 100 >>1$). The heavy neutrino abundance ($Y_{N_1}$), its equilibrium value ($Y^{\rm eq}_{N_1}$), and the generated lepton asymmetry ($Y_{\Delta L}$) are shown as functions of $z=M_1/T$. In this regime, efficient washout suppresses the final asymmetry but is largely independent of initial conditions.}
\label{fig:lepto-strong}
\end{figure}
\subsection{Strong and Weak Washout Regimes:}
Washout refers to the inverse decay process that depletes already created  lepton asymmetry. The washout effect is particularly important where heavy right handed neutrinos are not much heavier than the electroweak scale.The generated lepton asymmetry comes mainly from the out of equilibrium decays of heavy right handed neutrinos in type-I see saw. But the same neutrinos also include inverse decays and lepton number violating scattering which tends to erase the asymmetry. Whether the asymmetry survives depends strength of this washout processes which will be discussed below.

\begin{figure}[t!]
\centering
\includegraphics[width=0.89\textwidth]{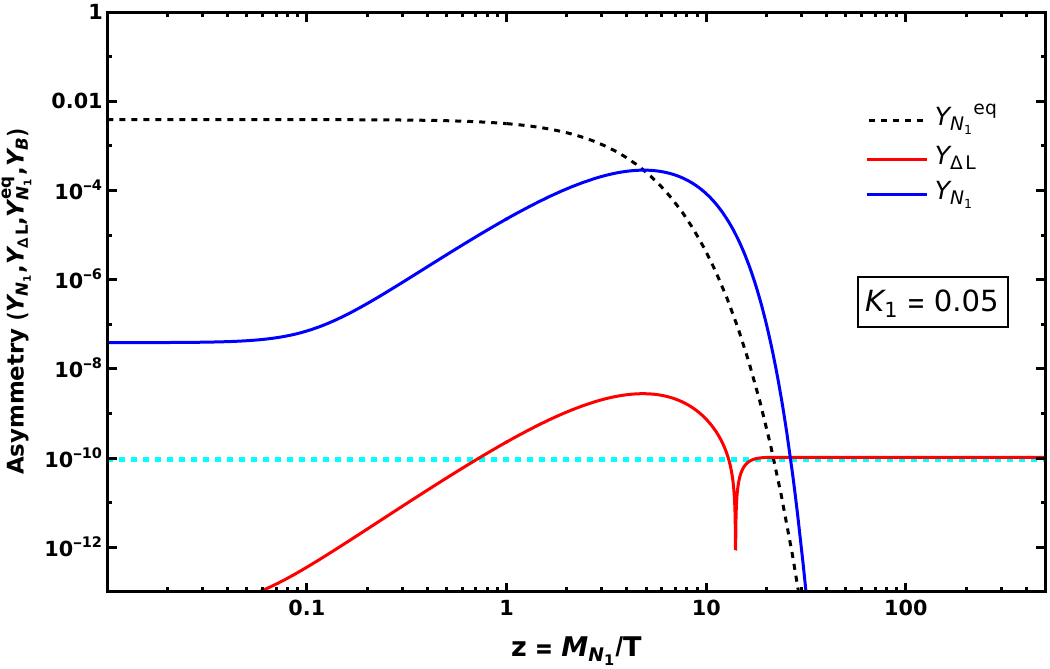}
\caption{Evolution plot of comoving number densities with the change of $z=T/M_1$ in the weak washout regime ($K_1 = 0.05 << 1$). Here, the washout processes are inefficient, allowing the lepton asymmetry to survive. The final asymmetry is partially depend on the initial conditions of the heavy neutrino abundance.}
\label{fig:lepto-wk}
\end{figure}

The evolution of lepton asymmetry and the final baryon asymmetry generation is numerically solved for two benchmark scenarios corresponding to the strong (as displayed in Fig. \ref{fig:lepto-strong}) and weak (presented in Fig. \ref{fig:lepto-wk}) washout regimes. 
We present the evolution of the comoving number densities in the strong washout regime with $K_1 = 100$ as displayed in Fig. \ref{fig:lepto-strong}. The blue line refers to the heavy neutrino abundance $Y_{N_1}$, the dashed black curve shows the equilibrium abundance $Y_{N_1}^{\text{eq}}$, and the red curve shows the generated lepton asymmetry $Y_{\Delta L}$. In the range of $ z\ll 1$, i.e $0.01 < z < 0.1$, the estimated values of the heavy RHN number densities $Y_{N_1}$ are small but trying to track its equilibrium value $Y_{N_1}^{\text{eq}}$. The dominant washout processes erase any pre-existing asymmetry in this region $0.01 < z < 0.1$. Thus,  the final baryon asymmetry is predominantly derived by the particular seesaw mechanism dynamics around freeze-out and is primarily independent of initial conditions. 

In the other hand, weak washout regime with $K_1=0.001-0.1 < 1$, the inverse and scattering processes are inefficient and as a result of this, heavy neutrino abundance deviates significantly from equilibrium as shown in Fig. \ref{fig:lepto-wk}. Quantitatively, the decay parameter is much smaller than unity, and the interaction rate is lower than the Hubble expansion rate. Thus, the heavy RHN abundance $Y_{N_1}$ does not initially track the equilibrium value $Y_{N_1}^{\text{eq}}$.  In this case, lepton asymmetry is generated earlier, and the washout processes are inefficient in erasing it, so the final asymmetry can be much larger compared to the strong washout case. Thus, the final lepton asymmetry partially depends on the assumed initial conditions of the right-handed neutrino population. In this regime, the initial condition on the $N_1$ density $(Y_{N_1}(z_i))$ is important for getting lepton asymmetry at $ z>> 1 $. 
Thus, both strong and weak washout regimes are complementary to each other, in which the strong washout ensures robustness of leptogenesis against uncertainties in the initial conditions, while the weak washout regime allows for larger surviving asymmetries but requires fine control over the initial heavy neutrino abundance. 

Electroweak sphalerons provide the essential bridge between lepton number violation at high scales and baryon number violation at the electroweak scale. The EW sphaleron process occurs at high energies, such as those present in the early universe shortly after the Big Bang. If there is an excess of particles carrying lepton number over antiparticles (a lepton asymmetry), then the EW sphaleron process could convert some of these particles into particles carrying baryon number, thus generating a baryon asymmetry. This is because the sphaleron can change the baryon and lepton numbers of the particles it interacts with, while still conserving the total amount of baryon and lepton number. If the lepton asymmetry generated through leptogenesis ends before the EW sphaleron process starts, the B-L asymmetry is expressed as $Y_{\Delta_{B-L}} =  - Y_{\Delta_{L}} $. At the later stage, when the EW sphaleron processes start, the B-L asymmetry is partially transferred to B asymmetry as:
\begin{equation}
    Y_{\Delta_{B}}(\infty)= \frac{28}{79}Y_{\Delta_{B-L}}(\infty)
\end{equation}

\section{Conclusion} 

We have studied a non-supersymmetric version of the Type I seesaw model in which the flavour structure is dictated by non-holomorphic modular forms of level three (polyharmonic Maass forms transforming under $\Gamma_3\simeq A_4$). The role of modular symmetry is to reduce the model parameters in the Yukawa sector by replacing conventional flavor fields with modular Yukawas. The structure of Yukawa couplings depends only on  the complex modulus $\tau$; thereby, the mass matrices are controlled by $\tau$ and a few other minimal parameters. The framework simultaneously explains the observed pattern of lepton mixing and provides calculable predictions for observables that connect low-energy neutrino physics to high-scale leptogenesis.

In the numerical analysis, we varied the real and imaginary part of the modular parameter, i.e, $\mathrm{Re}\,\tau,\mathrm{Im}\,\tau$ within the allowed range and scanned the other input model parameters, limited to the normal ordering of the light neutrinos. We examined two illustrative benchmark points for Dirac and Majoaran masses: 1) BP1 : Low  $M_D$ with $\tau=0.216+i\,0.996$, $\alpha_D\simeq 0.11,\beta_D\simeq 0.123,\gamma_D\simeq 0.083$, $\Lambda_1\simeq 3.70\times10^{12}$~GeV, $\Lambda_2\simeq 2.35\times 10^{13}$~GeV with estimated values of the light neutrino masses are $m_1\simeq 0.001$~eV, $m_2\simeq 0.009$~eV, $m_3\simeq 0.05$~eV and 
2)  BP2 : High  $M_D$ with $\tau=0.216+i\,0.997$, $\alpha_D\simeq 0.523,\beta_D\simeq 0.434,\gamma_D\simeq 0.58$, $\Lambda_1\simeq 8.69\times10^{13}$~GeV, $\Lambda_2\simeq 6.67\times10^{14}$~GeV yielding light neutrino masses $m_1\simeq 0.0014$~eV, $m_2\simeq 0.0089$~eV, $m_3\simeq 0.045$~eV.  Interestingly, these benchmark points are not only explain charged lepton masses, neutrino masses and mixing consistent with oscillation data but used for estimation of CP-asymmetry and final baryon asymmetry of the universe correctly. 

An important phenomenological implication of the present framework concerns neutrinoless double beta decay ($0\nu\beta\beta$), which serves as a key probe of the Majorana nature of neutrinos and lepton number violation. In our analysis, the estimated effective Majorana mass parameter $m_{ee}$ lies in the sub-meV to few-meV range, $m_{ee}\in [1.9\times 10^{-4},\,5.2\times 10^{-3}]$~eV, for normal ordering. This model predictions on effective Majorana mass parameter and half-lifes are well below the present experimental sensitivities from GERDA~\cite{GERDA:2020xhi}, KamLAND-Zen~\cite{KamLAND-Zen:2022tow}, and EXO-200~\cite{EXO-200:2019rkq} which constrain $m_{ee}\lesssim (0.1-0.4)$~eV as well as the forthcoming ton-scale experiments such as CUPID~\cite{CUPID:2019imh}, LEGEND-1000~\cite{LEGEND:2021bnm}, and nEXO~\cite{nEXO:2021ujk}, whose projected sensitivities extend down to $\mathcal{O}(0.006 - 0.027)$~eV. This covers the entire parameter space of $m_1 - m_{ee}$ for Inverted Ordering and a very thin space in the upper region of the Normal Ordering . However, the limit of $m_{ee}$ that we obtained from our model can be accessed in the future beyond the ton-scale experiments like ORIGIN-X \cite{Heffner}, THEIA\cite{Kaptanoglu}, LEGEND-6000\cite{LEGEND:6000} etc as shown in Fig. \ref{fig:expts}.  Moreover, the tight parameter correlations imposed by the modular symmetry ensure that any future measurement of $0\nu\beta\beta$ would place stringent tests on the viability of this model, especially for the normal ordering scenario where the allowed region is already highly constrained. 

In addition to low-energy predictions, the same modular Yukawa structures also provide the necessary ingredients for successful high scale thermal leptogenesis. In the framework of type-1 seesaw with $A_4$ modular symmetry the generation of the lepton asymmetry relies on the out-of-equilibrium decays of the heavy right-handed neutrinos. For successful leptogenesis, their decay rate must not be significantly larger than the Hubble expansion rate $\textbf{H}(T)$ of the Universe at the temperature $T = M_{N_1}$. 
The non-holomorphic $A_{4}$ modular symmetry determines both Dirac and Majorana mass textures, which in turn dictate the CP-violating phases and decay parameters of the right-handed neutrinos. We estimated the values of CP-asymmetry for both the benchmark points investigated and these values lie in the range $|\varepsilon_{1}|\sim 10^{-8}$--$10^{-4}$ consistent with expectations for RHN masses $M_{1}\sim 10^{9}$--$10^{14}$~GeV. Solving the Boltzmann equations~\cite{Buchmuller:2004nz,Davidson:2008bu} shows that the dynamics typically falls in the strong-washout regime, where the final baryon asymmetry is largely independent of the initial conditions.
Moreover, the generated baryon asymmetry values, consistent with the observed $\eta_{B}\simeq 6\times 10^{-10}$~\cite{Planck:2018vyg},  requires an efficiency factor $\kappa_f$ in the order of $10^{-3} - 10^{-5}$, which is compatible with the expected efficiencies in the strong-washout regime once flavor effects are included~\cite{Abada:2006ea, Blanchet:2006be}. Thus, the present $A_4$ modular seesaw framework with non-holomorphic modular forms naturally provides very concise and predictive paradigm that unifies neutrino masses, CP violation and baryogenesis for minimally fed model parameters.

\section*{Acknowledgments}
SP acknowledges the financial support under MTR/2023/000687 funded by SERB, Govt. of India.

\appendix 

\section{Appendix}
\subsection{Modular \texorpdfstring{$A_4$}{A4} invariant Lagrangian}

 The modular-invariant Yukawa Lagrangian for charged leptons and neutrinos is constructed by contracting field representations with modular multiplets so that both the representation and weight constraints are satisfied. The $A_4$ based modular-invariant Lagrangian relevant for the neutrino masses and mixing is written as
\begin{eqnarray}
&&-\mathcal{L}_{M_{A_{4}}}= \mathcal{L}_{M_\ell}+\mathcal{L}_{M_{D}}+\mathcal{L}_{M_R} \, , \nonumber \\
&& \mathcal{L}_{M_\ell}=\alpha_\ell Y^{(0)}_{3} \overline{L}_{L} H e_R 
+ \beta_\ell Y^{(0)}_{3} \overline{L}_{L} H \mu_R 
+ \gamma_\ell Y^{(0)}_{3} \overline{L}_{L} H \tau_R + {\rm h.c.} \nonumber \\
&&
  \mathcal{L}_{M_D} = \alpha_{D}\,(\overline{L_{L}} Y^{(-2)}_{3})_{1} N_{1R} H
+ \beta_{D}\,(\overline{L_{L}} Y^{(-2)}_{3})_{1^{\prime\prime}} N_{2R} H
+ \gamma_{D}\,(\overline{L_{L}} Y^{(-2)}_{3})_{1^{\prime}} N_{3R} H + {\rm h.c.} \nonumber \\
&&
  \mathcal{L}_{M_R} = \Lambda_1 Y^{(0)}_{1} N_{1R} N_{1R} 
  + \Lambda_2 Y^{(0)}_{1}(N_{2R} N_{3R} + N_{3R} N_{2R}) + {\rm h.c.}
\end{eqnarray}
where all interaction terms are \(A_4\)-invariant contraction into a singlet and the coefficients $\alpha_\ell, \beta_\ell, \gamma_\ell, \alpha_D, \beta_D, \gamma_D$ are free parameters (complex in general) to be determined phenomenologically. After electroweak symmetry breaking the Dirac and charged-lepton mass matrices take the explicit forms displayed in the main text Eq.(\ref{Eq:Md}) and Eq.(\ref{Eq:Me}) with each matrix element proportional to a modular form component $Y^{(k)}_{3,i}(\tau)$ multiplied by the corresponding prefactor $\alpha_\ell, \beta_\ell, \gamma_\ell, \alpha_D, \beta_D, \gamma_D$.
\begin{figure}
    \centering
    \includegraphics[width=0.5\linewidth]{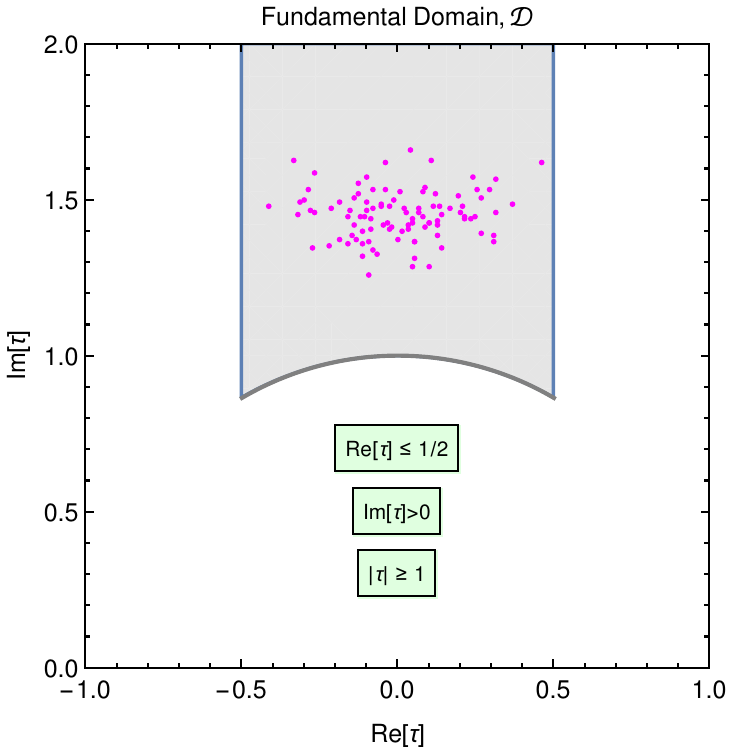}
    \caption{We have schematically shown the Fundamental Domain for any data points of $Re[\tau] \; \text{versus} \; Im[\tau]$ satisfying the constraints of a Fundamental Domain in the upper half plane $\mathcal{H}$ .} 
    \label{fig:placeholder}
\end{figure}

\subsection{Yukawa couplings with different modular weights}
\label{app:yukawas}
We present analytic expression for various truncated series values of various Yukawa couplings involved in the charged lepton and neutrino mass matrices used in our numerical analysis.  We follow the notation of  the complex modulus parameter $\tau$ and another relevant parameter $q$ as
\begin{eqnarray}
\tau = x + i y, \quad  x\equiv\operatorname{Re}[\tau], \quad y\equiv\operatorname{Im}[\tau],\quad q=e^{2\pi i\tau}.
\end{eqnarray}

The evaluated modular forms with different modular weights contributing Yukawa couplings are complex-valued functions of $\tau$. The analytic approximated expression for truncated Yukawa couplings are present below:

{\bf Weight $0$ triplet $\boldsymbol{Y^{(0)}_3(\tau)}$:}
\begin{eqnarray}
&&Y^{(0)}_{3,1}(\tau) = y - \frac{3 e^{-4\pi y}}{\pi q} - \frac{9 e^{-8\pi y}}{2\pi q^2} 
    - \frac{e^{-12\pi y}}{\pi q^3} - \frac{21 e^{-16\pi y}}{4\pi q^4} 
    - \frac{18 e^{-20\pi y}}{5\pi q^5} - \frac{3 e^{-24\pi y}}{2\pi q^6} + \dots \nonumber \\
    &&\hspace{2cm} - \frac{9 \log 3}{4\pi} - \frac{3q}{\pi} - \frac{9q^2}{2\pi} - \frac{q^3}{\pi} 
    - \frac{21q^4}{4\pi} - \frac{18q^5}{5\pi} - \frac{3q^6}{2\pi} + \dots, \nonumber \\
&&
    Y^{(0)}_{3,2}(\tau) = \frac{27 q^{1/3} e^{\pi y /3}}{\pi} \left( \frac{e^{-3\pi y}}{ 4q} 
    + \frac{e^{-7\pi y}} {5q^2} + \frac{5 e^{-11\pi y}} {16q^3} + \frac{2 e^{-15\pi y}} {11q^4} + \frac{2 e^{-19\pi y}} {7q^5 }
    + \frac{4 e^{-23\pi y}} {17q^6} + \dots \right) \nonumber \\
    &&\hspace{2cm} + \frac{9 q^{1/3}}{2\pi} \left( 1 + \frac{7q}{4} + \frac{8q^2} {7} + \frac{9q^3} {5} + \frac{14q^4} {13} 
    + \frac{31q^5} {16} + \frac{20q^6} {19} + \dots \right),  \\
&&
    Y^{(0)}_{3,3}(\tau) = \frac{9 q^{2/3} e^{2\pi y /3}}{2\pi} \left( \frac{e^{-2\pi y}}{ q} 
    + \frac{7 e^{-6\pi y}} {4q^2} + \frac{8 e^{-10\pi y}} {7q^3} + \frac{9 e^{-14\pi y}} {5q^4} 
    + \frac{14 e^{-18\pi y}} {13q^5} + \frac{31 e^{-22\pi y}} {16q^6} + \dots \right) \nonumber \\
    &&\hspace{2cm} + \frac{27q^{2/3}}{\pi} \left( \frac{1}{4} + \frac{q} {5} + \frac{5q^2} {16} + \frac{2q^3} {11} 
    + \frac{2q^4} {7} + \frac{9q^5} {17} + \frac{21q^6} {20} + \dots \right) \nonumber
    \label{Y03}
\end{eqnarray}

{\bf Weight $-2$ triplet $\boldsymbol{Y^{(-2)}_3(\tau)}$:}
\begin{eqnarray}
&&
    Y^{(-2)}_{3,1}(\tau) = \frac{y^3}{3} + \frac{21\Gamma(3,4\pi y)}{16\pi^3 q} 
    + \frac{189\Gamma(3,8\pi y)}{128\pi^3 q^2} 
    + \frac{169\Gamma(3,12\pi y)}{144\pi^3 q^3} 
    + \frac{1533\Gamma(3,16\pi y)}{1024\pi^3 q^4} + \dots \nonumber \\
    &&\hspace{2cm} + \frac{\pi}{40} \frac{\zeta(3)}{\zeta(4)} + \frac{21q}{8\pi^3} 
    + \frac{189q^2}{64\pi^3} + \frac{169q^3}{72\pi^3} 
    + \frac{1533q^4}{512\pi^3} + \frac{1323q^5}{500\pi^3} 
    + \frac{169q^6}{64\pi^3} + \dots \nonumber \\
 &&   Y^{(-2)}_{3,2}(\tau) = -\frac{729q^{1/3}}{16\pi^3} 
    \left( \frac{\Gamma(3,8\pi y/3)}{16q} 
    + \frac{7\Gamma(3,20\pi y/3)}{125q^2} 
    + \frac{65\Gamma(3,32\pi y/3)}{1024q^3} 
    + \frac{74\Gamma(3,44\pi y/3)}{1331q^4} + \dots \right) \nonumber \\
    &&\hspace{2cm} -\frac{81q^{1/3}}{16\pi^3} 
    \left( 1 + \frac{73q}{64} 
    + \frac{344q^2}{343} 
    + \frac{567q^3}{500} 
    + \frac{20198q^4}{2197} 
    + \frac{4681q^5}{4096} + \dots \right),    \\
&&
    Y^{(-2)}_{3,3}(\tau) = -\frac{81q^{2/3}}{32\pi^3} 
    \left( \frac{\Gamma(3,4\pi y/3)}{q} 
    + \frac{73\Gamma(3,16\pi y/3)}{64q^2} 
    + \frac{344\Gamma(3,28\pi y/3)}{343q^3} 
    + \frac{567\Gamma(3,40\pi y/3)}{500q^4} + \dots \right) \nonumber \\
     &&\hspace{2cm} -\frac{729q^{2/3}}{8\pi^3} 
    \left( \frac{1}{16} 
    + \frac{7q}{125} 
    + \frac{65q^2}{1024} 
    + \frac{74q^3}{1331} + \dots \right). \nonumber
     \label{Y-23}
\end{eqnarray}
We used $q=\exp(2\pi i\tau)$, $\tau$ and $y=\Im \tau$ to evaluate the incomplete Gamma function $\Gamma(s,x)$,  the Riemann zeta function $\zeta$ and thereby, all relevant Yukawa couplings. As a result, the charged-lepton and Dirac neutrino mass matrices are constructed using input model parameters and derived Yukawa couplings (as presented above) as 
\begin{eqnarray}
    M_\ell = v \, Y^{(0)}_{\ell}(\tau;\,\alpha_\ell,\beta_\ell,\gamma_\ell),\qquad
    M_D = v \, Y^{(-2)}_{D}(\tau;\,\alpha_D,\beta_D,\gamma_D),
    \label{ML-MD}
\end{eqnarray}
    where $v$ is the Higgs vacuum expectation value with $v\simeq 174\ \mathrm{GeV}$ and the complex prefactors $\alpha,\beta,\gamma$ are model parameters to be fitted for giving correct neutrino masses and mixings. Finally, one can diagonalize $M_\ell M_\ell^\dagger$ to obtain $U_\ell$, diagonalize $m_\nu=-M_D M_R^{-1}M_D^T$ to obtain $U_\nu$, and form $U_{\rm PMNS}=U_\ell^\dagger U_\nu$. 
\begin{table}[ht]
\centering
\small
\begin{tabular}{|c|c|c|c|c|}
\hline
 & $\tau=0.026 + i 1.714$ & $\tau=0.200 + i 1.200$ & $\tau=-0.300 + i0.900$ & $ \tau = 0.400 + i 1.800$  \\
\hline
$Y^{(0)}_{3,1}(\tau)$ & $0.93$ & $0.413 $ & $0.115$ & $1.013 $  \\
\hline
$Y^{(0)}_{3,2}(\tau)$ & $0.04 + i 002$ & $0.115 + i0.037$ & $0.190 - i0.082$ & $0.022 + i0.023$  \\
\hline
$Y^{(0)}_{3,3}(\tau)$ & $0.4 - i 0.002$ & $0.115 - i0.037$ & $0.190 + i0.082$ & $0.022 - i0.023$ \\
\hline
$Y^{(-2)}_{3,1}(\tau)$ & $1.77$ & $0.665 - i0.006$ & $0.323 + i0.021$ & $2.031 + i0.000$  \\
\hline
$Y^{(-2)}_{3,2}(\tau)$ & $-0.021 + i 0.002$ & $-0.062 + i0.050$ & $-0.066 - i0.134$ & $-0.001 + i0.010$  \\
\hline
$Y^{(-2)}_{3,3}(\tau)$ & $-0.153 + i0.008$ & $-0.226 + i0.101$ & $-0.229 - i0.176$ & $-0.093 + i0.103$  \\
\hline
\end{tabular}
\caption{Modular Yukawa triplets with three components evaluated at various representative complex moduli parameter $(\tau)$. These Yukawa couplings can be used to construct the charged-lepton and neutrino mass matrices by multiplying each triplet component by the corresponding complex prefactor (e.g. $(\alpha_\ell,\beta_\ell,\gamma_\ell,\alpha_D,\beta_D,\gamma_D)$) as described in the main text. The detailed discussion expansions listed in Appendix~\ref{app:yukawas}.}
\label{tab:modular-yukawas-tau}
\end{table}

We numerically estimated modular triplets Yukawa couplings in terms of various representative value of modular function $\tau$ which are presented in Table.\ref{tab:modular-yukawas-tau}. These illustrative benchmark points can be used to obtain the correct charged-lepton masses and fit the neutrino oscillation data while making the appropriate choice of the RHN mass matrix $M_R$.

\bibliographystyle{JHEP_improved}
\bibliography{./refsA4M}
\end{document}